# Leading edge bluntness effects on the hypersonic flow over the double-wedge at multiple aft-wedge angles

Anurag Adityanarayan Ray, [1)] Ashoke De, [1, 2, a)]

[1] *Department of Aerospace Engineering, Indian Institute of Technology, Kanpur, 208016, India*

[2] *Department of Sustainable Energy Engineering, Indian Institute of Technology, Kanpur, 208016, India*

The present numerical investigation focuses on the leading edge bluntness effects on the double-wedge with varied aft-wedge angles exposed to low enthalpy hypersonic free stream conditions. The bluntness ratio in this study varies, ranging from $R/L_1 = 0$ (sharp leading edge) to $R/L_1 = 0.577$ (maximum allowable bluntness), along with the aft-wedge angle varying between $\theta_2 = 45^0$ and $60^0$. Noticeably, even a small bluntness ratio can completely change the shock interaction pattern compared to its sharp geometrical counterpart due to a detached leading edge shock, enlarged separation bubble, and location of various shock-waves concerning it. Critical bluntness ratios exist for the low aft-wedge $\theta_2 = 45^0$ angle, but increasing the aft-wedge angle makes the flow field highly unsteady for some bluntness ratios. Nevertheless, these bluntness ratios for such double-wedge configurations are reported using the mean of separation bubble size. Moreover, this work unravels the cause of such unsteadiness for the unsteady flow-fields using the spatial-temporal evolution of the wall pressure distribution, Fast Fourier Transform (*FFT*) of the pressure fluctuation signal at the compression corner and supports the deduced observation with the help of energy-based Proper Orthogonal Decomposition (*POD*). The increased shock-boundary layer interaction strength moves the separation point upstream beyond the junction of cylindrical bluntness and inclined fore-wedge surface, accompanying sudden change in its direction of motion that perturb the shear layer that set to a self-sustained, highly unsteady flow field.

___________________________

[a)] Author to whom correspondence should be addressed: ashoke@iitk.ac.in

## I. INTRODUCTION

Shock interference and shock-wave boundary layer interaction are ubiquitous in the internal and external hypersonic flows. In the case of the external flows, interactions of the shocks originating out of the various components of a hypersonic vehicle can cause severe fluctuating pressure loads and high heat transfer rates, which may result in its catastrophic failure. This was precisely the cause of the failure of the structural integrity of the hypersonic cruise vehicle X-15[1]. Its airframe was attached with a dummy ramjet engine on the bottom of the fuselage through a pylon. During its flight, the unrevealed interaction of the shock waves led to the burnout of the fuselage's bottom and the loss of the ramjet engine. Numerous efforts have been made in the past few decades to unravel the mystery behind this interaction. B. Edney performed a successful experimental investigation to understand the physics behind the shock interference patterns[2] by impinging the planar oblique shock wave onto the bow shock at various locations, creating 6 different interference patterns, and classifying them based on the classical shock-polar diagrams. The author also measured wall pressure and heat flux distribution over the cylinder's surface, correlated them to these interference patterns, and concluded that the type-IV interaction was the most severe, which generated highly unsteady pressure and heat transfer loads. In later years, a similar interference phenomenon was observed on canonical configurations such as double wedges. Olejniczak *et al.*[3] studied experimentally and numerically the behavior of the shock waves in the presence of a boundary layer on the double-wedge configuration exposed to hypersonic flow conditions. They observed discrepancies in the experimental and numerical results and owed it to the inefficient modeling of the thermal and chemical non-equilibrium. It thus motivated them to analyze the interference patterns on the double-wedge. Olejniczak *et al.*[4] studied the shock-interference phenomenon on the double-wedge configurations in the varied parametric space of the aft-wedge angles at a free-stream Mach Number $(M_\infty) = 9$. They found similar interference patterns in their inviscid numerical simulations aided by the shock-polar diagrams, which were earlier reported in experimental observations of Edney but with an exception. They reported a new interference pattern at one of the aft-wedge angle and classified it as 'type-IVr', where an under-expanded supersonic jet glides on the surface of the wedge to create oscillating pressure distribution. Moreover, they reported only Edney type-I, IV, V, or VI shock interference over geometrically constrained flows. This computational investigation provided new insights into the physics associated with such a complicated flow field, and hence, extensive experiments and computational and analytical studies were performed by several researchers.

Hu *et al.*[5] performed numerical computations to analyze the unsteady interaction of the shock waves on the double-wedge configuration. They performed inviscid two-dimensional computation, but in contrast to the previous study by Olejniczak *et al.* they used multi-species, temperature-dependent thermo-physical properties to model the free-stream air. They found multiple self-sustained oscillation mechanism with different shock-interference shapes, which transits from 7-shock to 7-shock, 7-shock to 6-shock as well as 7-shock to 6-shock and again back to 7-shock system transition at varied free-stream Mach numbers and aft-wedge angles. They even claimed that such oscillatory transition occurred at higher aft-wedge angles than its perfect gas simulation counterpart. Ben-Dor et al.[6] instead conducted numerical simulations for the aft-wedge angle in the range $\theta_2 = 35^0$ and $\theta_2 = 50^0$ and back to the initial aft-wedge angle, which they varied with time. They reported that the flow field undergoes a transition in shock-interference shape from regular to Mach reflection for the change in the difference between aft-wedge and for-wedge angles ($\Delta\theta = \theta_2-\theta_1$) for $26.4^0$ and $28^0$, respectively. However, the flow field maintains a self-sustained oscillation across these two states for the narrow range of $\Delta\theta = 26.8^0$ onwards up to $28^0$, where the flow field settles down at steady-state. Seshadri and De[7] investigated another transient supersonic flow over stationary and moving wedge. The investigated geometry had similar geometrical parameters, except that the second angle is convex instead of concave. The authors used a sharp interface-based immersed boundary method with a fifth-order weighted essentially non-oscillatory scheme and varied the parametric space of the second angle ($60^0$, $90^0$, and $120^0$) and incident shock Mach numbers (1.3, 1.9, and 2.5). They analyzed the mechanism of vorticity generation and claimed that the generation is highly dependent on the viscous effects compared to the baroclinic or compressible effects.

Such interesting observations reported across the computational fluid dynamics community provided an imminent interest among the experimentalists to verify such interaction process. Swantek and Austin[8] carried out one such experimental investigation. They fabricated a double-wedge with fixed fore-wedge length ($L_1$) and aft-wedge length ($L_2$) of *50.8 mm* and *25.4 mm,* respectively *($L_1/L_2 = 2$)* with span-wise length $L_z = 101.8$ mm. Instead of varying the geometrical configuration, they varied the free-stream conditions at different Mach numbers *4-7* and Reynolds number *0.435-4.64 x 10$^6$/m* in conjunction with two different media, which were air and pure Nitrogen ($N_2$). The authors reported their observations for the short run time of the experiments using the schlieren flow-visualization technique and complemented their results with the heat transfer profiles at mid-span and offset of mid-span. They reported Edney type-V interaction for all the conditions, but the difference between the bow shock standoff distances grows significantly with the increase in free stream enthalpy for the air and Nitrogen case. Interestingly, the flow field is laminar for the free stream conditions with the highest and lowest enthalpy, irrespective of the test gas. The authors attributed this observation to the relatively low Reynolds number in the free stream. Hashimoto[9] performed similar experiments considering three different experimental models with the fixed fore-wedge angle but different aft wedge angled $\theta_2 = 40^0$, $50^0$, and $68^0$ and reported the observations in the form of flow-visualization based on interferogram and unsteady schlieren images. The author observed Edney type VI and type V interference patterns in the visualization for the aft-wedge angle $\theta_2 = 40^0$ and $\theta_2 = 50^0$, respectively. A strong type-V interaction was also reported for $\theta_2 = 68^0$, where the strong impinging shock-wave induced complex unsteadiness in the flow field.

These experimental observations lead to an interesting development in understanding the physics behind these interaction processes. However, the experimental setups have very low run time (in order of microseconds), and at such a short run time, it is difficult to ascertain whether the flow reaches a steady state. Hence, different research groups from the SATO-AVT-205 task group conducted an extensive study to compare their numerics against the experimental data provided by Swantek and Austin[10]. They proposed various numerical strategies based on the Direct simulation Monte Carlo *(DSMC)* and several types of Navier Stokes solvers with the different capabilities of the numerical schemes using the three-dimensional and two-dimensional assumptions and were able to establish a good agreement with the experimental results except near the peak heat transfer location (at the re-attachment point). These observations reincarnated interest among the computational fluid dynamics community to understand the cause of such inconsistency and to address the physics associated with the long run time durations.

In this context of the double-wedge flow, Komives *et al.*[11] performed a computational study for the geometry provided by Swantek and Austin at low flow enthalpy. They used numerical methods with first and second-order schemes to simulate and validate their flow solver. However, they found a good match with the experimental results at the separation and re-attachment region but failed to predict correctly in the well-attached region upstream of the interaction owing to the high amount of artificial viscosity inherited in low-order schemes. In addition, the flow field sets into the self-sustained periodic state at large flow-through times for higher-order discretization. The authors concluded that one must be careful when selecting the numerical discretization schemes based on these observations. This investigation was extended further with an entire three-dimensional flow field, compared with the two-dimensional flow, and reported differences at various locations. It was attributed to the side-ways relaxation of the flow due to low span-wise length, which induces span-wise vortices. However, their three-dimensional results did not match, especially near the peak heat transfer location. Durna *et al.*[12] also studied the physics associated with such interactions over the double-wedge on the same double-wedge configuration but only at low freestream enthalpy conditions using $N_2$ as the test gas. In their two-dimensional viscous interactions, they varied the aft-wedge angle in the range of $\theta_2 = 45^0$ and $60^0$ systematically and reported that the flow transits from a weak type V interaction at a lower aft-wedge angle to strong type V interaction at a high aft-wedge angle in their short run-time duration study. They reported a transition angle at the aft-wedge angle between $\theta_2 = 45^0$ and $50^0$. Durna *et al.*[13] published an article that discussed in depth the behavior of the flow field exposed to low enthalpy hypersonic free-stream conditions for large flow through time. Their two-dimensional analysis revealed that the flow field undergoes self-sustained oscillations for a threshold value of the aft-wedge angle greater than $\theta_2 = 47^0$. Further, they correlated the transmitted shock's duration with its strength and suggested that the impingement period shortens at higher aft-wedge angles. Durna *et al.*[14] extended their study on double-wedge flow by simulating the unsteady flow using two-dimensional and three-dimensional computations for the short flow-through time (corresponding to the experimental steady state time reported by Swantek and Austin[8]). They observed that, during the initial shock-establishment time, there were no significant differences between the two and three-dimensional flow fields; however, substantial differences were observed beyond it. The authors reasoned these observations to the side-wise relaxation of the flow field, which led to the small size of the separation bubble in contrast to the two-dimensional flow field. Additionally, they observed stream-wise striations in the three-dimensional wall heat-flux distribution, indicating the presence of Görtler vortices, which became prominent at higher aft-wedge angles, making the flow field highly asymmetric about the center line. Nevertheless, the three-dimensional study still did not provide the correct location of the peak heat transfer rate.

Kumar and De[15] further used this numerical investigation campaign to address the inconsistency between the computational and experimental results. The authors deployed an in-house solver, which they claimed to be highly stable up to Courant number (*CFL*) 1. They used a preconditioner to solve the governing equations simultaneously in OpenFOAM© instead of sequential integration, which increased the default solver *"rhoCentralFoam"* stability. With their systematic grid, time independence study, and the appropriate time averaging window, they could predict the mean wall-heat flux distribution over the double-wedge's surface, which entirely agrees with the experimental results. Moreover, they extended their study further at varied aft-wedge angles and aft-wedge length ratios for extended run time duration and observed oscillatory flow field for the double-wedge configurations with longer aft-wedge length compared to the baseline geometry. The authors reasoned it to the impingement of the transmitted shock wave close to the geometry's expansion corner, which alternately causes contraction and relaxation of the separated bubble. Kumar and De[16] extended their work further by varying the parametric space of the aft-wedge angle in the range of *$45^0$-$60^0$* and the aft-wedge length ratio *$L_1/L_2$* in the range of *0.5 to 2*. They observed that the flow field behaves according to the imposed geometrical constraints and categorized these flow field as 'steady', 'pulsating', 'oscillatory', and 'vibrating' analogous to the spiked-body configurations[17]. These observations were attributed to the separation bubble's size variation and the shock structure's location.

Hao *et al.*[18] performed a numerical investigation over the same double wedge flow with hypervelocity conditions but with higher free stream enthalpy conditions. The authors performed time accurate two and three-dimensional simulations using pure Nitrogen and air as the test gas and found the shock structures are qualitatively in good agreement with the experiment result of Swantek and Austin. They predicted the mean heat flux close to the experimental values for the pure Nitrogen flow with two-dimensional results. However, they under-predicted the peak slightly in their three-dimensional results owing to expansion waves centered at the corner. The shock structures were similar for the high enthalpy air flow to the Nitrogen flow. Still, the peak heat flux was significantly lower than the experimental values. They attributed this discrepancy to the major contribution of high-temperature non-equilibrium effects in the air flow. Expósito et al.[19] also conducted a numerical investigation over the double wedge geometry for both low and high enthalpy free stream conditions. They used a non-equilibrium flow solver, 'hy2Foam' developed in OpenFOAM©[20]. The authors validated their numerical results for low enthalpy Nitrogen flow by comparing mean heat flux between $t = 50\ \mu s$ and $t = 327\ \mu s$ against the experimental data. They could obtain a good match within the experimental uncertainty except near the peak heat flux location, which was substantially over-predicted. They further reported the flow field to be highly unsteady within the experimental run time and did not find any significant dissociation of the Nitrogen molecule at low enthalpy and reasoned this observation to its high amount of bond dissociation energy.

Some computational studies also focused on reproducing the high enthalpy experimental results using numerics e.g. Jiang and Yan[21] simulated the same geometry at high enthalpy by considering both laminar and turbulent flow assumption by modeling the gas as a calorically perfect and thermally perfect gas. They reported the best match with the experimental data by assuming the flow to be turbulent and the gas model as thermally perfect. On the contrary, Dai et al.[22] numerically investigated the same case of high enthalpy with air by using four different modeling of the gas vis-à-vis thermochemical non-equilibrium gas *(TCNEG)*, thermally non-equilibrium chemical freezing gas *(TNCFG)*, chemical non-equilibrium gas *(CNEG)*, thermal equilibrium without chemical reactions *(TPG)*. The authors also considered the impact of laminar and turbulent flow assumptions. Their simulations reported the highest instantaneous heat flux ($t = 170\ \mu s$) by modeling the gas as TPG within the laminar regime. In contrast, the CNEG model predicted the highest heat flux in the turbulent flow regime. Such differences among various models were an artifact of the different sizes of interaction regions, hence the impinging shock's strength. The TCNEG model with laminar flow assumption best predicted the experimental data qualitatively and quantitatively. Ninni et al.[23] recently simulated the same double wedge flow using low and high enthalpy flow. The authors considered pure Nitrogen for low enthalpy-free stream conditions, whereas Nitrogen and air were used for high enthalpy simulations. Their results showed subtle differences concerning the low and high enthalpy at large flow through time. The flow field was periodic for low enthalpy free stream conditions while it was completely steady at high enthalpy conditions, irrespective of the test gas.

Vatanesever and Celik[24] predicted the wall heat flux distribution over the double wedge surface at high enthalpy free stream conditions using their solver named *'hyperReactingFoam'* in the OpenFOAM environment. They could predict the mean wall heat flux distribution over the wedge's surface except at the peak location, where they substantially under predicted. On the contrary, they obtained a good match of the instantaneous heat flux distribution with the previously established numerical results. Moreover, the authors considered four different aft-wedge angles in the range $\theta_2 = 45^0$ and $\theta_2 = 60^0$ with the fixed fore-wedge to aft-wedge length ratio $L_1/L_2 = 2$. They reported severity in the heat flux measurement and enhanced fluctuations with the increase in the aft-wedge angle. Further, the reaction rates accelerated at high aft-wedge angles. Another interesting numerical study by Hong et al.[25] of the flow over a double wedge exposed to high free stream enthalpy provided exciting insights into various transition angles using different modeling of the test gas. The authors considered three models corresponding to frozen flow, thermochemical non-equilibrium flow, and thermal non-equilibrium. They reported the delay of critical aft-wedge angles for the incipient separation angle, the appearance of the secondary vortices, and the transition of the interference pattern by including vibrational relaxation and chemical dissociation as the translational mode's energy redistributes among the vibrational and chemical internal energy mode.

The previous studies mentioned above were based on sharp double-wedge configurations. Still, the sharp configuration is practically avoided, which would otherwise lead to high heat-transfer rates and pressure loads at the leading edge. The impact of the leading edge bluntness on the hypersonic flow over various canonical configurations was studied. Holden[26] first analyzed the impact of leading edge bluntness with a few different bluntness radii and reported that the size of the separation bubble for the case of double-cone geometry is highly dependent on it. The large size of the separation bubble was reported for the high bluntness radius; however, the flow field is marginally affected by a small bluntness radius. The cause of this interesting observation was unclear till then. In similar lines, Holden[27] experimentally and analytically accessed the impact of the leading edge bluntness effects on the size of the separation bubble using the flat-plate wedge junction and noticed that the separation bubble size increases and then declines with the increase in bluntness ratio. John and Kulkarni[28] in their computational study, reported similar observations based on two-dimensional laminar flow assumption. They pointed out two essential bluntness ratios; one corresponding to the maximum size of the separation bubble and termed it inversion radius, and the other corresponding to the equivalent size of the separation bubble observed on its sharp leading edge counterpart and termed it equivalent radius. They correlated it to the relative dominance of the entropy and hydrodynamic boundary layers. For bluntness ratios lesser than the critical radius, the hydrodynamic boundary layer 'swallows' the entropy layer, but beyond the inversion radius, the entropy layer overtakes the hydrodynamic boundary layer. The numerical study by Khraibut et al.[29] showed that the separation bubble size grows considerably at high leading edge bluntness, which even promoted the appearance of a secondary separation bubble beneath it. A recent numerical study by Hao and Wen[30] using the spherically blunted double cones exposed to low and high enthalpy flows revealed the existence of an inversion radius and equivalent radius for the double cone configuration. The authors correlated the strength of the shock-interference pattern to the bluntness ratios, where the shock interference shape largely remains unaltered at a very low bluntness ratio but is highly disgorged at high bluntness ratios with reduced shock intensity. These observations were similar irrespective of the free-stream enthalpy. Exposito et al.[31] performed a three-dimensional numerical simulation for the SWBLI interaction over the flat plate/wedge surface by considering the sharp and blunt leading edge with different wall temperatures. The authors reported the flow field to be two-dimensional at the symmetry plane, and three-dimensional effects were primarily confined to the edges of the geometry. Moreover, they observed the Edney type VI interference pattern with two triple points. The distance between them reduces with the increase in the wall temperature and eventually collapses into a single triple point at the highest wall temperature ($T_w = 1080\ K$). This observation pertains to the separation bubble size's increment with the increasing wall temperature. The authors further reported incrementing the separation bubble's size by including the leading edge bluntness.

Many previous studies cited above-considered analysis of the double-wedge configuration with a sharp leading edge; however, such leading edges are generally blunt, as it is either reinforced intentionally to mitigate the adversities of the SBLI or introduced due to machining inaccuracy while manufacturing, in reality. Although the studies on leading edge bluntness provided exciting insights into the flow field, most focused on either weak SBLI (type VI) or flow featuring a small separation bubble size (double-cone). Previous studies cited above on the sharp double wedge reported the transition of the interference pattern for the aft-wedge angle between $\theta_2 = 45^0$ and $\theta_2 = 50^0$. The authors believe that the influence of leading edge bluntness is rarely explored in the transition and strong SBLI interaction regime that can create a large separation bubble comparable to the characteristic dimension of the geometry. Hence, this article comprehensively analyzes the cylindrical leading edge bluntness with multiple aft-wedge angles of the double wedge.

This work considers the angle's variation between $\theta_2 = 45^0$ and $\theta_2 = 60^0$, representing weak and severe interaction, respectively, whereas the bluntness ratio varies between $R/L_1 = 0$ (sharp leading edge) and $R/L_1 = 0.577$ (maximum bluntness ratio based on the fixed fore-wedge angle, $\theta_1 = 30^0$) to accomplish the objectives of the present study.

The outcome of this article will help to analyze the cylindrical leading edge's impact on the double-wedge configuration, which is essentially the building block of a hypersonic intake. It will shed light on various aspects of regimes of flow representing steady and unsteady flow-fields at a sufficiently large flow-through time and additionally report the influence of bluntness on the separation bubble's size. These results will ultimately help choose an appropriate leading edge bluntness for such hypersonic intakes, which, if not chosen correctly, may bring in additional instabilities and, therefore, can cause the engine intake to unstart[32-36].

The rest of this article is organized as follows. Section II provides the necessary governing equations based on continuum flow assumptions. Section III discusses the mathematical implementation and algorithm of the solver employed in the present study. Section IV shows the details of the dimensions of the geometrical configuration along with the relevant information on the computational domain and boundary conditions. Section V is about the grid configuration used in this numerical investigation and subsequent validation of the solver. Section VI is about the results obtained for the blunted double-wedge configuration at four different aft-wedge angles. Section VII summarises the results obtained in section VI, and the article concludes with Section VIII with the prominent results and the possibility of further extension of this study.

## II. GOVERNING EQUATIONS

The flow field considered in the present study is laminar and compressible. The conventional unsteady Navier-Stokes equation in conservative form governs this flow field and is shown below[37-40]:

$$\frac{\partial}{\partial t}\iiint_\Omega \vec{W}d\Omega + \iint_{\partial\Omega}(\vec{F_c}-\vec{F_v})dS = \iiint_\Omega \vec{Q}d\Omega \tag{1}$$

$$\vec{W} = \begin{pmatrix} \rho \\ \rho\vec{u} \\ \rho E \end{pmatrix} \tag{2}$$

$$\vec{F_c} = \begin{pmatrix} \rho(\vec{u}.n) \\ \rho\vec{u}(\vec{u}.n)+pn \\ (\rho E+p)(\vec{u}.n) \end{pmatrix} \tag{3}$$

$$\vec{F_v} = \begin{pmatrix} 0 \\ \overline{\overline{\tau}}.n \\ (\overline{\overline{\tau}}.n).\vec{u}+\kappa(\nabla T).n \end{pmatrix} \tag{4}$$

$$\vec{Q} = \begin{pmatrix} 0 \\ \rho\vec{f} \\ \rho\vec{f}.\vec{u}+\dot{q}_h \end{pmatrix} \tag{5}$$

Here, $\vec{W}$ represents the vector of the conserved variables, $\vec{F_c}$ corresponds to the inviscid flux vector, $\vec{F_v}$ is the viscous flux vector, and $\vec{Q}$ is the source term vector. $\rho$ is the density of the fluid, $\vec{u}$ is the flow velocity vector, $E = e + 0.5\rho|\vec{u}|^2$ is the total energy of the fluid comprising of the internal energy $e = C_v.T$ and flow kinetic energy $0.5\rho|\vec{u}|^2$. Here $C_v$ is the specific heat capacity of the fluid at constant volume. The thermodynamic pressure is represented by $p$, the viscous shear stress tensor is given by $\overline{\overline{\tau}} = \mu_v\left(\nabla\vec{u}+(\nabla\vec{u})^T-\frac{2I}{3}\nabla.\vec{u}\right)$ ($I$ is the unit stress tensor of second order) and $T$ represents temperature. Heat generation within the fluid element is $\dot{q}_h$ and $n$ is the unit normal vector pointing outwards (away from the control volume). The fluid is modeled as a Newtonian, with its dynamic viscosity $\mu_v$ formulated using Sutherland's law. Fourier's law of conduction models the heat flux vector wherein the thermal conductivity of the medium is related through the Prandtl number (*Pr*). The flow medium is a thermally perfect gas (*TPG*) that models the high-temperature effects of vibrational energy excitation by accounting for it in the expression of specific heat at constant volume ($C_v$). It is a polynomial function of temperature only whose constants are provided with the help of *JANAF* tables[41]. Finally, ideal gas law $p = \rho RT$ provides closure to the above set of governing equations. The flow is inherently laminar, as reported in the experiments and a few of the computational studies[11-16]; hence, turbulence modeling is omitted in the present study.

Conservation equations are solved numerically by employing the finite volume method (*FVM*) in the open-source platform OpenFOAM© [25] using the density-based solver with a semi-discrete, collocated, Gudonov-type solver. It uses the second-order central-upwind scheme of Kurganov-Noelle-Petrova (*KNP*)[42] for the convective flux discretization. Further, the slope limiter function based on the Total Variation Diminishing (*TVD*) reconstruction strategy facilitates interpolating the conservative variables onto the face center from the neighboring cell-centered data[43]. The solution to the governing equation advances in the time coordinate employing four-stage third-order accurate Runge-Kutta (RK-4) explicit time-integration. This solver is named *rhoCentralhypersonicRK4Foam*, an up-scaled version of the default operator-splitting algorithm-based solver *rhoCentralFoam*. It provides an added advantage regarding stability, scalability, and accuracy for Mach numbers ranging from subsonic to supersonic flows[44]. Finally, the viscous flux terms are discretized using a second-order accurate central scheme in space. The mathematical implementation of these discretization techniques is described in the next section.

## III. NUMERICAL FORMULATION AND SOLVER ALGORITHM

This section deals with the mathematical formulation of the discretization schemes for the contributing terms of the Navier-Stokes equation described previously. The above conservation equation can be expressed in the form of general conservation law as[45],

$$\frac{\partial W}{\partial t} + \nabla . F(W) = 0 \qquad (6)$$

Where $W = \{W_1, W_2, W_3......W_n\}$ represents the set of conserved variables in the appropriate form and $F(W)$ is the flux function. The conservation equation can be further evaluated using numerical strategy and the relevant boundary condition. Finite Volume Method *(FVM)* based formulation is used in the present investigation to convert it into a set of algebraic equations which can be solved simultaneously to seek the solution of the conservation law. The *FVM* based formulation uses the principle of integration of these conservation laws over an arbitrary control volume *(CV)*. Integrating Eq. 6 over an arbitrary control volume $\Omega_i$ with volume magnitude $|\Omega_i|$ gives

$$\frac{d\overline{W}}{dt} = -\frac{1}{|\Omega_i|} \iiint_{\Omega_i} F(W) dV \qquad (7)$$

where, $\overline{W}$ represents the cell averaged conserved quantity over the *CV*. Eq. 7 is further simplified by invoking the Gauss Theorem, which transforms the volume integration on the right hand side of this equation into a surface integral over $\partial S$, which is the magnitude of the surface area of the *CV* $\Omega_i$. The surface of the *CV* is assumed to contain a finite number of faces for simplicity; hence, the final expression reduces to

$$\iiint_{\Omega_i} F(W) dV = \oiint_{\partial S} F(W).dS \approx \sum_f F(W_f).S_f \qquad (8)$$

here $S_f$ is the face normal vector of any arbitrary face $\Omega_i$.

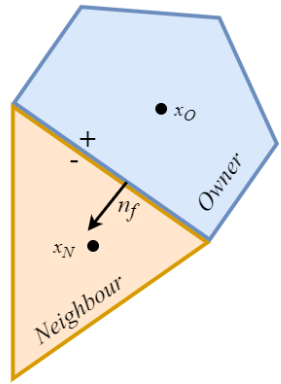

FIG .1. Typical control volumes used in the finite volume formulation in OpenFOAM©, along with the definition of + and – sides of the control surface.

### A. Numerical fluxes

The flux function in Eq. 8 contains the contribution of convective and viscous fluxes. The viscous fluxes can be evaluated using standard linear/weighted interpolation because of their diffusive nature, while on the other hand, the computation of the convective fluxes is demanding. This sub-section of the article briefly glances through the numerical formulation of the convective flux formulation. The physical convective (inviscid) fluxes $F(W)$ are split numerically into two parts, *i.e.* $\Psi(W_f^+, W_f^-)$ to account for the hyperbolic nature of the conservation equations in high speed flows[37, 45]. Note the $\pm$ refers to the owner and neighbor cells, respectively, as shown in Figure 1. The OpenFOAM© package computes the normal vector of the cell face center such that it points in the direction from the owner to the neighbor side (Figure 1). Hence, it is apparent that $W_f^+$ the owner biased interpolation $W_f^-$ is the neighbor cell's contribution to the face center. Estimating these fluxes in the present study employs the Kurganov-Noelle-Petrova[42] scheme, a blend of central-upwind based formulation. The split numerical flux for this convective scheme is presented below

$$\Psi^{KNP}(W^+, W^-) = \frac{\lambda^+ F(W^+) - \lambda^- F(W^-)}{\lambda^+ - \lambda^-} + (\frac{\lambda^+ \lambda^-}{\lambda^+ - \lambda^-})[W^- + W^+] \tag{9}$$

$\lambda^+$ and $\lambda^-$ are the fastest disturbance propagation speed (wave speed) in the direction normal to the face. Note $\lambda^+$ carries the disturbance towards the neighbor cell (along the face normal), whereas $\lambda^-$ carries the disturbances in the opposite direction, *i.e.* towards the owner cell. Their computation is the crux of the algorithm, and they are formulated as

$$\lambda^+ = \max\{(u^+.n_f + c^+), (u^-.n_f + c^-), 0\}$$
$$\lambda^- = \max\{(u^+.n_f - c^+), (u^-.n_f - c^-), 0\} \tag{10}$$

where, $u$ is the local velocity vector of the fluid, $c = \sqrt{\gamma RT}$ is the local speed of sound in the medium, $R$ is the characteristic gas constant of the fluid, and $n_f$ denotes the unit face normal vector.

## B. Flux reconstruction

The flux computations on the face centers were shown in the preceding section, but the computation of the flow variables from both the sides of the face (owner and neighbor cells) of the control volume onto the face center needs to be performed to compute them. The help of TVD based interpolation schemes can achieve such owner and neighbour biased interpolation. These schemes must be robust and accurate to compute the flow with strong shocks and other discontinuities. In this article, van Leer formulation is also suggested in Ref. 46 to have the least dissipation. This section briefly deals with the scalar variable interpolation only, but the detailed formulation of the vector variable is available in Ref. 46. The *TVD* interpolation of the scalar variable is defined as

$$W_f^+ = \overline{W_O} + (1-\omega_f)\zeta(r_{f,O})(\overline{W_N} - \overline{W_O})$$
$$W_f^- = \overline{W_N} + \omega_f\zeta(r_{f,D})(\overline{W_O} - \overline{W_N})$$
(11)

here, $\overline{W_O}$ and $\overline{W_N}$ represents the cell averages of any scalar variable in the owner and neighbor cells, respectively. The factor $r_{f,i}$ provides the measure of the smoothness of the solution[46], $\omega_f$ and $(1-\omega_f)$ provides the relative weights/contribution of the scalar variable based on the relative cell size[46]. The $r_{f,i}$ is calculated based on the expression shown below irrespective of structured and unstructured mesh[46]. $(\nabla W)_i$ represents the gradient of the scalar variable computed at any corresponding cell $i$ and $d_{ON} \equiv x_N - x_O$ is the vector pointing from the owner cell towards the neighbor cell.

$$r_{f,i} = \frac{2(\nabla W)_i \cdot d_{ON}}{\overline{W_N} - \overline{W_O}} - 1 \qquad \text{where } i = O, N \tag{12}$$

The smoothness factor $r_{f,i}$ is used further to compute the slope limiter function $\zeta(r_{f,i})$ based on *TVD* formulation. The following expression is used for the van Leer interpolation

$$\zeta_{vanLeer}(r_{f,i}) = \frac{|r_{f,i}| + r_{f,i}}{1 + r_{f,i}} \tag{13}$$

**C. Temporal integration**

The previous section briefly described the spatial discretization of various terms appearing on the right hand side of Eq. 6. This section will glance through the time integration strategy adopted in this numerical study. It was found previously by Kumar and De[15] that the default operator based splitting approach of the default *rhoCentralFoam* solver may cause unphysical spurious oscillation in the numerical solution, especially behind the strong shock waves. They formulated their integration strategy as fully implicit time integration and used the preconditioner to have stability in all the flow regimes (incompressible to supersonic). The new solver was found to be stable up to *CFL* 1. The implicit formulation needs linear solvers to find the solution to the system of algebraic equations. These computations can be pretty expensive if the solution is sought for the large-scale problems[44] and thus can offset the advantage of the fast rapid convergence of the solver. Hence, this study employs explicit integration with 4 stage 3rd order accurate Runge-Kutta (RK4) time integration strategy. The scalability analysis of the solver is already performed by Sibo *et al.*[44] and proved the scalability and accuracy of the RK-4 based time integration; therefore, the present study omits this aspect.

Eq. 6 is residual based formulation[37,] where the residual of this equation is computed using the initial and boundary conditions and is subsequently evolved in time through the numerical time integration performed between time $t_n$ and $t_{n+1}$. This numerical methodology can be expressed mathematically as

$$\overline{W}^l = \overline{W}^0 + \alpha_l \Delta t \left[ -\frac{1}{|\Omega_i|} \oiiint_{\Omega_i} F(\overline{W}^{l-1}) dV \right], \qquad l = 1,2,3......N_s \qquad (14)$$

here, $\overline{W}^0 \equiv \overline{W}(t_n)$ and $\overline{W}^{N_s} \equiv \overline{W}(t_{n+1})$, $\alpha_l$ are the corresponding weights in sub-stages of the integration, $\Delta t$ is the time-step increment, and $N_s$ is the number of sub-stages of the integration (calculated based on stability conditions of the solver based on *CFL* number and grid Fourier number). For the *RK-4* strategy, the number of sub-stages equals *4,* and corresponding weights are set as $\alpha_1 = 0.25, \alpha_2 = 1/3, \alpha_3 = 0.5, \alpha_4 = 1.0$.

### D. Solution technique

The previous section briefly illustrates the discretization of all the terms that appear in the conventional Navier-Stokes equation. This section describes the solution methodology and algorithm adopted in the present study. The unsteady equations are solved sequentially in the order of appearance, starting with the mass conservation equation and ending with the total energy conservation equation. The pressure field is thus evaluated at the end of the integration via the aid of the ideal gas law assumption. Algorithm 1 below shows the pseudo-code of the present solver for clarity.

---
*Algorithm 1: Unsteady solver 'rhoCentralhypersonicRK4Foam'*

**Start:** time loop for unsteady integration
    Evaluate $\rho, (\rho\vec{u}), (\rho E)$ from previous time step into corresponding variables
    **Start**: RK4 loop: (RK4 sub-stages)
        Interpolate fluxes on the face using reconstruction schemes
        Evaluate convective and viscous fluxes for mass, momentum, and energy equation
        Solve density, momentum, and energy equations using boundary conditions
        Calculate the pressure field using the ideal gas law
    **End** (end RK4 sub-stage)
    Update time step using stability criteria
**End** (end time loop)

---

## IV. COMPUTATIONAL DOMAIN AND BOUNDARY CONDITIONS

Figure 2(a) shows the computational domain used in the present study. The double wedge configuration has a fixed fore-wedge length of $L_1$ = 50.8 mm *and* an aft-wedge length of $L_2$ = 25.4 mm. The fixed fore-wedge and aft-wedge angles concerning the freestream directions are $\theta_1 = 30^0$ and $\theta_2$, varying in the range of $45^0$ and $60^0$, respectively. The computational domain is constructed to contain the region close to the Shock-boundary-layer-interaction (SBLI) zone to assess the impact of bluntness on the separation bubble size. Point A represents the intersection of the symmetry plane and the inlet patch, which is $L_1/8$ distance upstream of the apex of the double wedge.

Further, the domain is extended horizontally beyond the expansion corner by $0.75L_1$ to avoid the reflection of outwards propagating information back into the domain. Finally, the inlet patch intersects the outlet patch at point B above the expansion corner, placed vertically at $1.5L_1$ from it. To validate the flow parameters, the aft-wedge angle is set at $55^0$, and the bluntness ratio $R/L_1 = 0$, which mimics Swantek's experimental geometric configuration as shown in Figure 2(b)[47]. The configuration without a blunt leading edge will be called a 'sharp double-wedge' in the upcoming sections. The leading edge is blunted with the bluntness addition strategy mentioned in[30], except that it is cylindrical instead of spherical. Note the maximum allowable bluntness ratio restricted for this double-wedge geometry corresponds to $R/L_1 = 0.577$, and the effective fore-wedge length shortens by adding bluntness to the geometric using this method. The x-coordinate direction is set along the wedge axis, whereas the y-coordinate direction points perpendicular to it.

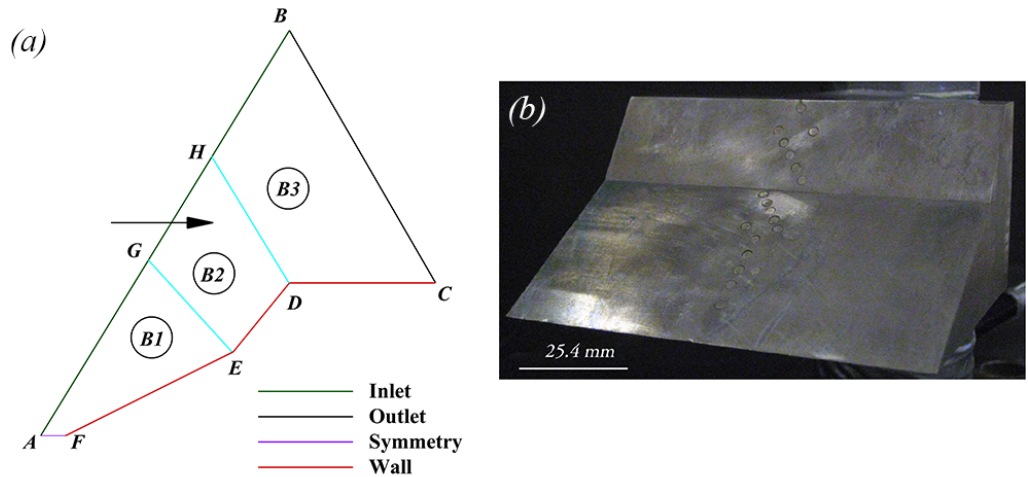

FIG .2. Geometrical details of double wedge configuration (a) computational domain and (b) experimental model [Reproduced with permission from A. Swantek, "The Role of Aerothermochemistry in Double Cone and Double Wedge Flows." Ph.D. thesis (University of Illinois at Urbana-Champaign, 2012). [47].]

In the present two-dimensional analysis, the inlet of the domain is the uniform free-stream boundary with $U_\infty = 1972$ m/s, $T_\infty = 191$ K, and $p_\infty = 391$ Pa (Dirichlet condition), which corresponds to the low enthalpy, free-stream conditions used by Swantek *et al.* in their experimental study[8]. The domain's exit must be non-reflective; the wave-transmissive boundary condition is applied by solving one dimensional linearized wave equation ($\frac{D\varphi}{Dt} \approx \frac{\partial \varphi}{\partial t} + U_n \cdot \frac{\partial \varphi}{\partial n} = 0$) perpendicular to the outlet-patch.

Here, $\varphi$ is any variable crossing the domain's exit, $U_n = u_n + c$ is the wave speed along the outlet face vector's normal direction, $u_n$ is the local velocity vector oriented along the face vector's direction, and $c$ is the local speed of sound. The wall is isothermally maintained along with no-slip boundary conditions concerning both velocity ($\vec{u_w} = 0$) and temperature ($T_w = 300$ K), replicating the experimental conditions with the Neumann condition for pressure gradient ($\frac{\partial p}{\partial n} = 0$). The symmetry boundary condition ($\frac{\partial p}{\partial n} = 0, \frac{\partial T}{\partial n} = 0, \frac{\partial v_t}{\partial n} = 0, \frac{\partial v_n}{\partial n} \approx 0, v_n = 0$ here $v_t$ and $v_n$ represent parallel and perpendicular to the symmetry plane velocity components) is used for the patch upstream to the leading edge (*LE*). Note $n$ represents the boundary's normal direction. The initial conditions for this numerical investigation inside the domain are free-stream velocity ($U_\infty$), pressure ($p_\infty$), and temperature ($T_\infty$). The test gas in the present numerical study is pure di-atomic Nitrogen ($N_2$).

## V. GRID INDEPENDENCE TEST AND VALIDATION

The computational domain is discretized with the block-structured grid, as shown in Figure 2, and blocks are labeled (B1) to (B3). To optimize the computational resources and the accuracy of the solution, the grid is prepared arbitrarily from a certain refinement level. It then systematically improves by incrementing the number of the grid points both in stream-wise and cross-stream directions. Through this process, three distinct grids of increasing refinement levels are generated and labeled as 'coarse', 'medium', and 'fine'. The medium grid is refined by a factor of *1.5* concerning the coarse grid, and the fine grid is *4* times the number of grid points in a coarse grid in all directions. The exact block edge matching information for all the refinements is given in Table I referring to the double wedge with zero bluntness ratio. Note the subscripts on the spacing in different directions shows the name of the connecting vertices shown in Figure 2, where the edge is directed from the first indexed vertex to the second (e.g. FE is the first cell spacing on edge directed from F towards E and vice-versa) while the subscript *'w'* represents spacing in cross-stream direction. Table II provides the details regarding the number of grid points for the final converged mesh used in this numerical study. Note, for every grid in this study, the first cell height perpendicular to the wall direction is fixed at *5.08 μm* as suggested in Ref. 15 and is stretched gradually using geometric expansion law away from the wall with an expansion ratio of *1.01*. The cell spacing in a stream-wise direction is such that the aspect ratio is within an acceptable level for ensuring the stability of the solver (maximum aspect ratio < *560*). Figure 3 depicts the fine grid generated using the commercial grid generator ICEM-CFD® employing the above strategy, along with the locations of various refinement regions such as the compression corner, the leading edge region, and the trailing expansion corner. The grid spacing at various block line intersection zones matches exactly to minimize any errors in the region of strong gradients.

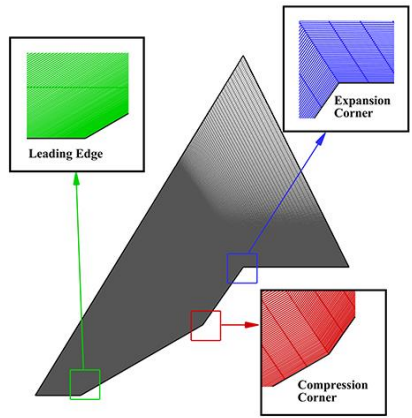

FIG. 3. Schematic of the fine grid configuration generating using a block-structured approach along with magnified images at various critical locations

TABLE I. First cell grid for different grid sizes spacing in stream-wise ($\Delta x$) and cross-stream directions ($\Delta y$) in $\mu m$.

| Grid | $\Delta x_{FE}$ | $\Delta x_{EF}$ | $\Delta x_{ED}$ | $\Delta x_{DE}$ | $\Delta x_{DC}$ | $\Delta y_w$ |
|---|---|---|---|---|---|---|
| Coarse | 500 | 500 | 500 | 500 | 500 | 5.08 |
| Medium | 400 | 400 | 400 | 400 | 400 | 5.08 |
| **Fine** | **100** | **100** | **100** | **100** | **100** | **5.08** |

TABLE II. Final converged grid size used for the analysis of different bluntness ratios

| $R/L_1$ range | $Nx$ | $Ny$ | Grid points |
|---|---|---|---|
| 0 | 838 | 512 | 4,29,056 |
| 0.05-0.25 | 1372 | 512 | 7,02,464 |
| 0.35-0.577 | 460 | 512 | 2,35,520 |

Figure 4 compares the numerical schlieren obtained at $t = 150\ \mu s$ for all three grids along with the experimental schlieren captured by Swantek and Austin[47]. The leading edge shock (*LeS*) and the bow shock (*BS*) intersect at a triple point *(TP)* to create a transmitted shock (*TS*) which impinges on the aft wedge surface and reflects off it to create a reflected shock *(RS)* along with the thickening of the boundary layer at the impingement point. This interaction's schematic diagram is available in Ref. 15, featuring the shock interaction mechanism along with the prominent details of the flow. Qualitatively, all the numerical schlieren generated from these grids provide reasonably unvaried locations of the *LeS*, separation shock (*SS*), bow shock *(BS)*, triple point *(TP),* and *TS*. Moreover, the size of the separation bubble is marginally different, but, significant differences are observed in the resolution of the shock wave discontinuities, slip lines emanating from the triple point, and shear layer originating from the separation point, which is highly dissipated for the coarse grid whereas they are sharply resolved for the fine grid resolution.

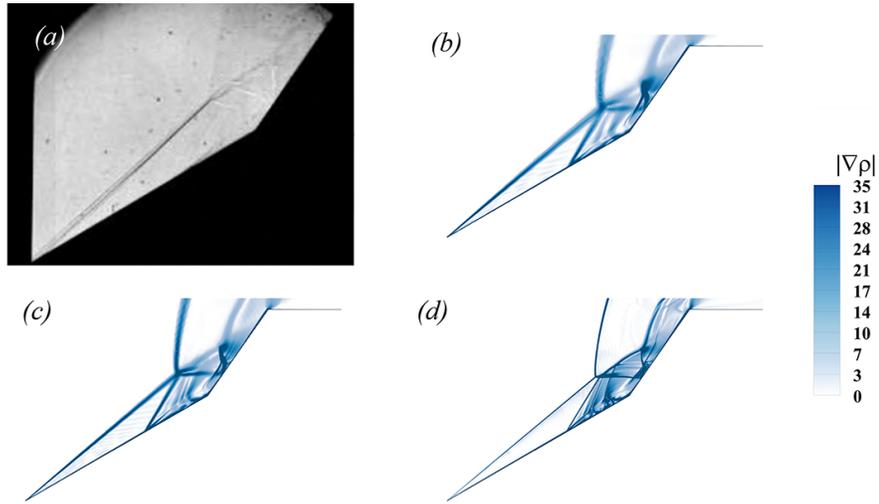

FIG. 4. Instantaneous density gradient magnitude ($kg/m^4$) at $t = 150\,\mu s$ using three different grids for $\theta_2 = 55^0$ and $L_1/L_2 = 2$ (a) Experimental schlieren. [Reproduced with permission from A. Swantek, "The Role of Aerothermochemistry in Double Cone and Double Wedge Flows." Ph.D. thesis (University of Illinois at Urbana-Champaign, 2012). [47].] (b) Coarse grid. (c) Medium grid. (d) Fine grid.

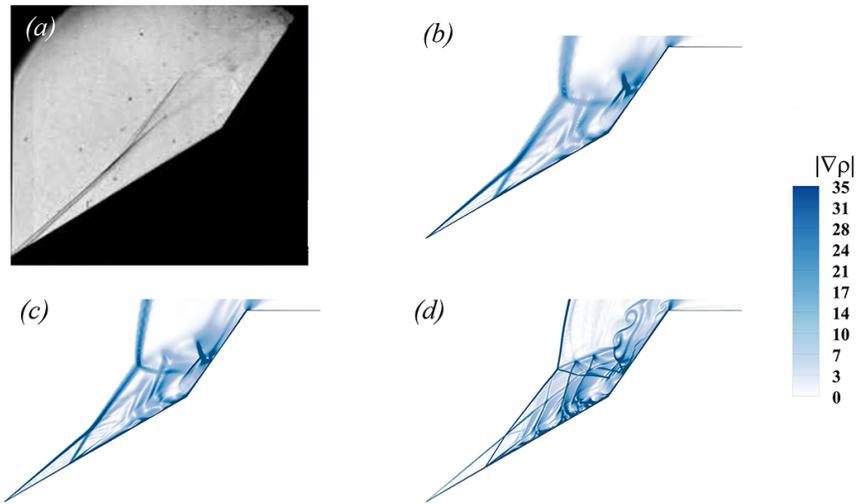

FIG. 5. Instantaneous density gradient magnitude ($kg/m^4$) at $t = 260\,\mu s$ obtained using three different grids for $\theta_2 = 55^0$ and $L_1/L_2 = 2$ (a) Experimental schlieren. [Reproduced with permission from A. Swantek, "The Role of Aerothermochemistry in Double Cone and Double Wedge Flows." Ph.D. thesis (University of Illinois at Urbana-Champaign, 2012). [47].] (b) Coarse grid. (c) Medium grid. (d) Fine grid.

The present simulation is highly unsteady[15] as the separation bubble size is still growing under the influence of the impinging transmitted shock. Hence, this qualitative comparison of grid refinement is extended beyond $t = 150\,\mu s$. Figure 5 presents the numerical schlieren at $t = 260\,\mu s$, depicting the substantial difference between the location of separation and transmitted shock impingement location forming a large separation bubble *(SB)*. The separation point *(SP)* moves upstream, which leads to the intersection of the *LeS* and the *SS,* to form an intermediate shock *(IS)*. The *IS*, the bow shock *BS* in front of the aft wedge, and the transmitted shock form the *TP*. The reader can refer to the detailed schematic diagram for this interaction mechanism at this time instance from Ref. 15. Figure 6 additionally reports the span-wise vorticity contour plot at the same time step $t = 260~\mu s$, near the compression corner of the double wedge junction. The subsequent refinement of the grid resolves the chain of vortices evident in Figure 6 (b) and more clearly in Figure 6 (c), which is one of the primary causes for the modes of unsteadiness observed with the hypersonic flow over double wedge configuration[16].

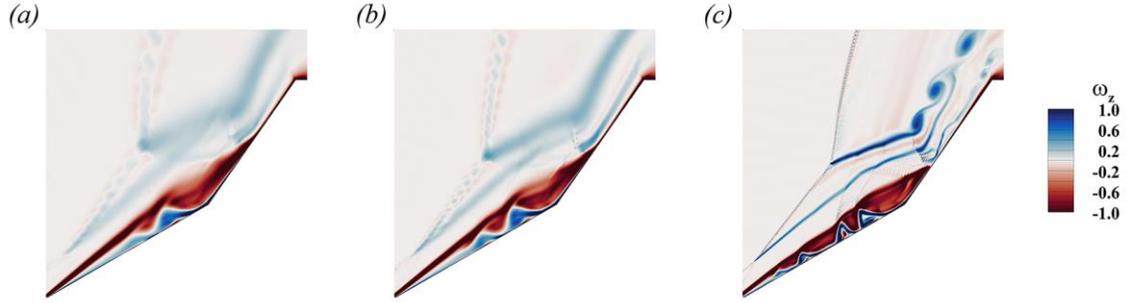

FIG. 6. Instantaneous z-vorticity, $\omega_z$ component (in million $s^{-1}$) at $t = 260~\mu s$ using the three different grids for $\theta_2 = 55^0$, $L_1/L_2 = 2$ and $R/L_1 = 0$ (a) Coarse grid. (b) Medium grid. (c) Fine grid.

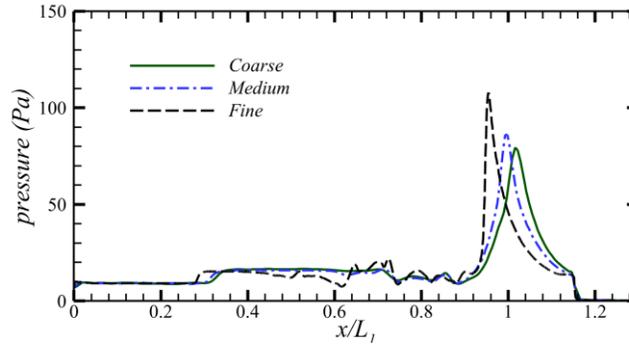

FIG. 7. Wall pressure distribution using three different grids for $\theta_2 = 55^0$, $L_1/L_2 = 2$, and $R/L_1 = 0$ at $t = 260~\mu s$.

The fine grid provides the sharpest resolution of shock structures and the vortical structures beneath the primary separation bubble compelling it to use for further analysis, but a lucid quantitative comparison is required to understand the behavior of grid refinement on the flow properties such as wall pressure and wall heat flux. Figure 7 illustrates the wall pressure profile at $t = 260~\mu s$. In this figure, both the *SP*, which is represented by the first rise in pressure due to flow compression by the *SS* and the re-attachment point *(RP)* corresponding to the location of impingement of the transmitted shock, move upstream upon refining the grid from coarse to the medium grid and they shift on further refinement. These observations are consistent with the findings reported in Ref. 15. The peak pressure also becomes distinct with the subsequent grid refinement with better resolution of transmitted shock, which is in mutual agreement with the observations in Figure 4 and Figure 5.

The investigation further addresses the influence of different grid resolutions on the wall heat flux. Many authors found the flow field highly unsteady in the numerical investigation, unlike the experimental observation after a large flow-through time[11-16]. Different time-averaging windows were used to obtain a reasonable match with the experimental data. Durna and Celik[12] suggested using a short time averaging window between 150 $\mu s$ - 310 $\mu s$ based on their theory of shock establishment. Following their strategy, the same time averaging windows are used for the present study and are compared with the experimentally acquired heat flux data. Figure 8 (a) reports the average heat flux obtained using the grid mentioned above resolutions in this short time window. The numerically obtained average wall heat flux data are in excellent agreement with the experimentally acquired profile, but a significant discrepancy is seen at the peak heat flux location. This is ascribed to the ambiguous time window reported in the experimental study and was confirmed by Kumar and De[15]. They suggested using a wider time average window, i.e. from the start of the simulation and ending at $t = 400$ $\mu s$ to improve the agreement with the experimental data. Following their theory, the average wall heat flux is estimated in this time window and compared with the experimental results shown in Figure 8 (b). A better match is seen with the experimental results than those obtained using a short time averaging window. The peak heat flux keeps improving along with its upstream movement, similar to the results shown in Figure 8 (a). It is not astonishing that the coarse grid and medium grid's results almost overlap with the experimental results owing to the very high dissipation introduced by the grid spacing, but such an unresolved grid may either introduce artificial oscillation in the flow field due to incorrect shock wave location or may dissipate the tertiary vortices responsible for the 'vibrating mode'[15,16]. Therefore, the fine grid best predicts the average heat flux in this wider time-averaging window.

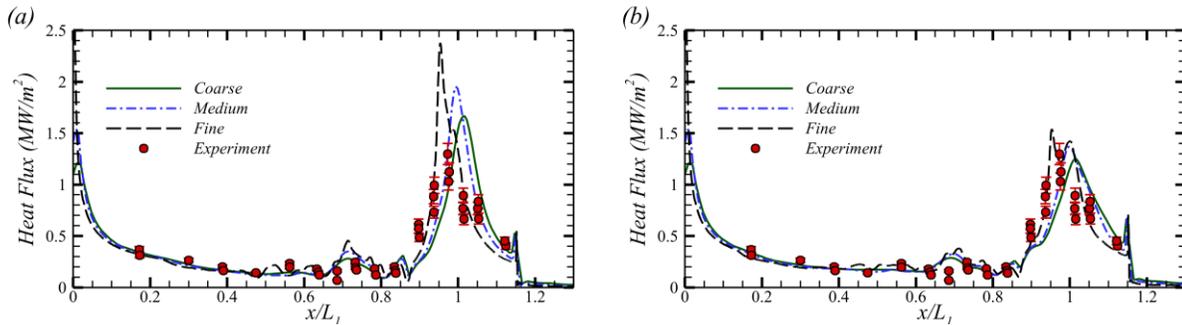

FIG. 8. Average wall heat flux distribution for $\theta_2 = 55^0$, $L_1/L_2 = 2$ and $R/L_1 = 0$ (a) Averaging time window between $t = 150$ $\mu s$ and 310 $\mu s$ (b) Averaging time window between $t = 0$ $\mu s$ and 400 $\mu s$. Error bars correspond to 8% uncertainty in experimental data. (Experimental data reproduced with permission from A. Swantek and J. Austin in 50$^{th}$ AIAA Aerospace Sciences Meeting Including the New Horizons Forum and Aerospace Exposition (2012), p. 284.[7] Copyright 2012 Andrew B. Swantek.)

Similarly, a separate grid independence test allows choosing the best grid size for the blunted double wedges using similar arguments suggested by John and Kulkarni[28] (the results are not shown here to maintain brevity). Table II shows the details of the converged grid topology for the geometries with different bluntness ratios to resolve the flow near the double-wedge's wall.

Therefore, the fine grid resolution (in Table II) corresponds to the bluntness ratios in this investigation to determine the impact of leading edge bluntness ratios.

## VI. RESULTS AND DISCUSSION

This section discusses the effects of bluntness ratios on different aft-wedge angles. The leading edge bluntness ratios vary in the range $0 \leq R/L_1 \leq 0.577$ for each of the aft-wedge angles in the range $45^0 \leq \theta_2 \leq 60^0$ and maintaining the fore-wedge angle ($\theta_1=30^0$) and aft-wedge length $L_2 =25.4$ mm constant. This range of aft-wedge angle covers the various regimes of the interference patterns ranging from weak type VI interaction at the lowest aft-wedge angle to the strong type V interaction at the highest wedge angle, which can create a large separation bubble at the corner. It further focuses on estimating the inversion and equivalent bluntness ratios corresponding to the maximum size of the separation bubble and equivalent bubble size, respectively. During this investigation, it is found that the flow field switches to a new-unsteady mode at higher aft-wedge angles. This new unsteady mode is addressed based on the spectral analysis, *i.e.* Fast Fourier Transform (*FFT*) of pressure signals and energy-based Proper Orthogonal Decomposition (*POD*).

### A. Aft-wedge angle $\theta_2 = 45^0$

Figure 9 reports the numerical schlieren at steady state for different bluntness ratios and its spatial-temporal evolution for wall pressure (Figure 10). The spatial-temporal plots for all these bluntness ratio combinations show that the flow field reaches a steady state at large flow through time. The shock-interference undergoes a weak type-V interaction for sharp-edged configuration, which was also reported by Durna *et al.*[12]. The separation and leading edge shock wave (*SS & LeS*) coalesce to form an intermediate shock (*IS*) which further meets the separation shock wave (*SS*) and the bow shock at the triple point *(TP)* forming a very weak transmitted shock *(TS)*. Moreover, the *TS* impinges far downstream of the sharp expansion corner *(EC)*, does not disrupt the boundary layer, and restricts the growth of the separation bubble, making the flow field steady. But in contrast, the shock interference pattern differs for the blunted configurations. Figure 9 (b) depicts the interaction corresponding to $R/L_1 = 0.05$, in which the separation point *(SP)* moves upstream greatly compared to the sharp leading edge geometrical configuration. The *LeS* is detached at the leading edge of the body. It interacts with the *SS* originating at the point of separation, forming *IS*. The *IS, RS,* and the Bow Shock *(BS)* meet at the third triple point to form an Edney-type VI interaction pattern and, as a result, produces a slip line along with a centered expansion fan, but it could not be resolved in the numerical schlieren as they are very weak. Moreover, there are multiple secondary separations beneath the primary separation bubble, but they do not produce any mode of unsteadiness. The *TS* is absent in this type-VI interaction which was identified as the primary source of unsteadiness for the flow over the double-wedge[13]. Furthermore, the shock-interference pattern weakens severely with an increase in the bluntness ratios, and for considerable bluntness ratios, e.g. $R/L_1 = 0.577$, forms a strong detached bow-shock in front of it, evident in Figure 9 (i). The shock waves such as *SS* and *RS* are absent and form a small separation bubble at the juncture of the two wedge planes for this case.

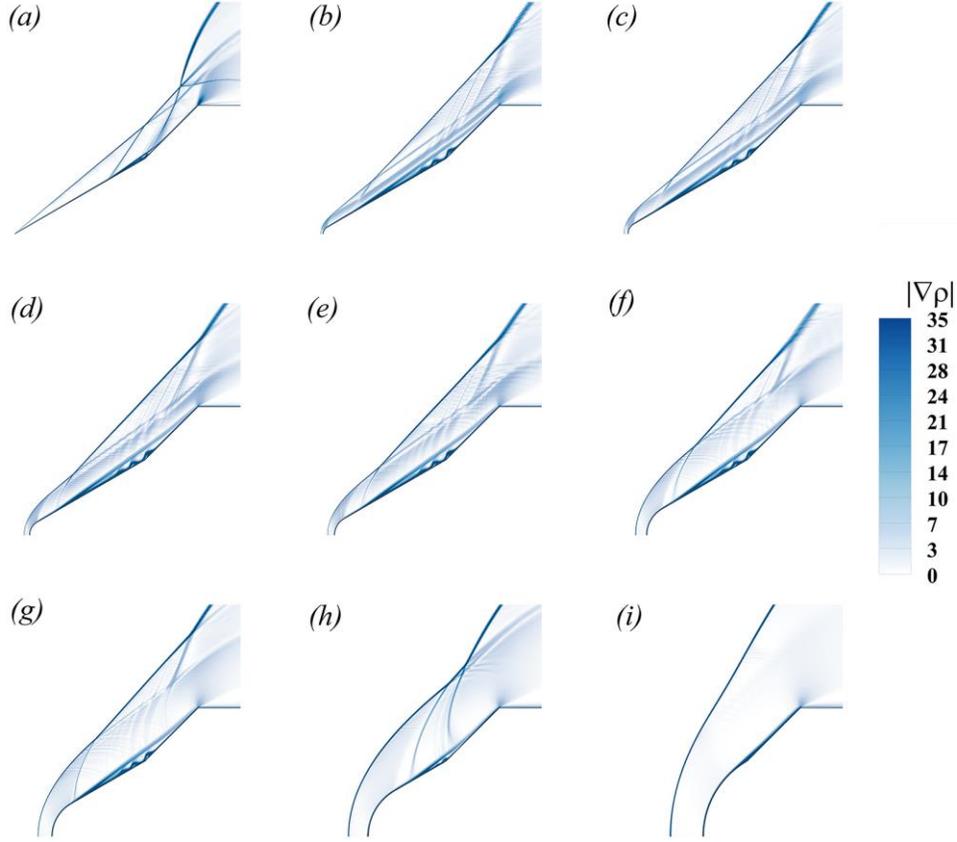

FIG. 9. Instantaneous numerical schlieren (density gradient magnitude in $kg/m^4$) at steady-state for $\theta_2 = 45^0$, $L_1/L_2 = 2$ (a) $R/L_1 = 0$, (b) $R/L_1 = 0.050$, (c) $R/L_1 = 0.075$, (d) $R/L_1 = 0.1$, (e) $R/L_1 = 0.125$, (f) $R/L_1 = 0.2$, (g) $R/L_1 = 0.25$, (h) $R/L_1 = 0.35$, (i) $R/L_1 = 0.577$.

Figure 11 illustrates the pressure variation over the double-wedge surface for different steady-state bluntness ratios (at $t = 2.5\ ms$). As expected, the pressure distribution profile is exorbitantly different for the blunted geometries compared to the sharp geometrical configuration. It falls off gradually from the stagnation pressure corresponding to the Rayleigh Pitot tube formula calculations due to the expansion waves over the blunted leading edge configurations. In contrast, the pressure that rises on the leading edge for the sharp configuration is due to an attached oblique shock wave. The pressure further rises downstream of the leading edge for all of these cases, indicating the separation location of the boundary layer. The pressure distribution fluctuates beyond this point due to multiple vortices present in the primary separation bubble and attains the maximum value for bluntness ratio in the range $0.05 \leq R/L_1 \leq 0.35$. The maximum pressure for the blunted geometries reduces significantly compared to sharp leading edge configuration indicating the weakening of the interaction and remains approximately constant for all the bluntness ratios except for the case of $R/L_1 = 0.577$ in which pressure rises by strong bow shock-wave (see Figure 9(i) and Figure 10(i)).

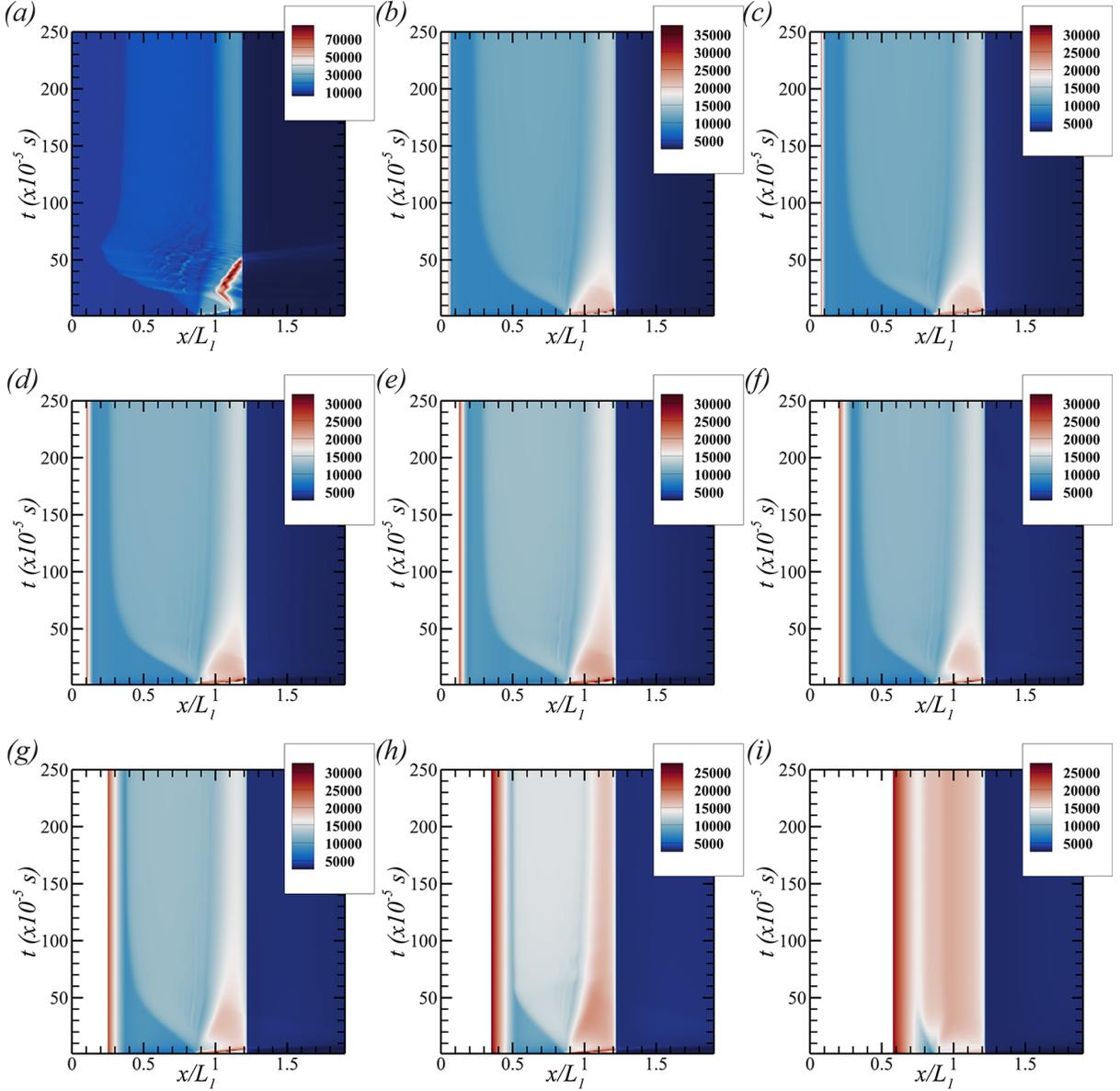

FIG. 10. Spatial-temporal evolution of wall pressure distribution for (a) $R/L_1 = 0$, (b) $R/L_1 = 0.050$, (c) $R/L_1 = 0.075$, (d) $R/L_1 = 0.1$, (e) $R/L_1 = 0.125$, (f) $R/L_1 = 0.2$, (g) $R/L_1 = 0.25$, (h) $R/L_1 = 0.35$, (i) $R/L_1 = 0.577$

Figure 12 demonstrates the variation in skin friction coefficient defined as $C_f = \tau_w/0.5\rho_\infty U_\infty^2$ for varied bluntness ratios at steady state. These plots quantify the size of the separation bubble by providing the accurate location of the separation and the reattachment point. They are readily obtained by determining the locations where the skin friction coefficients cross the zero line marked with a black dashed horizontal dotted line, as suggested by Hao and Wen[30]. Note the skin friction crosses this line multiple times, which confirms the presence of multiple separations beneath the primary bubble and is also reported earlier in this article using the numerical schlieren and the pressure distribution. Also, the skin friction falls rapidly downstream of the leading edge due to the expansion of the flow over the blunted nose. These skin friction distributions downstream of the leading edge are below the distribution obtained on sharp double-wedge configuration due to pressure over expansion[30]. This is the prime reason for the increase in the separation bubble size compared to the sharp double wedge configuration and is the consequence of the entropy layer swallowing by the boundary layer[28]. These plots also suggest that the length of the separation bubble increases till $R/L_1 = 0.075$ and then declines, with the size of the separation bubble being minimum for $R/L_1 = 0.577$.

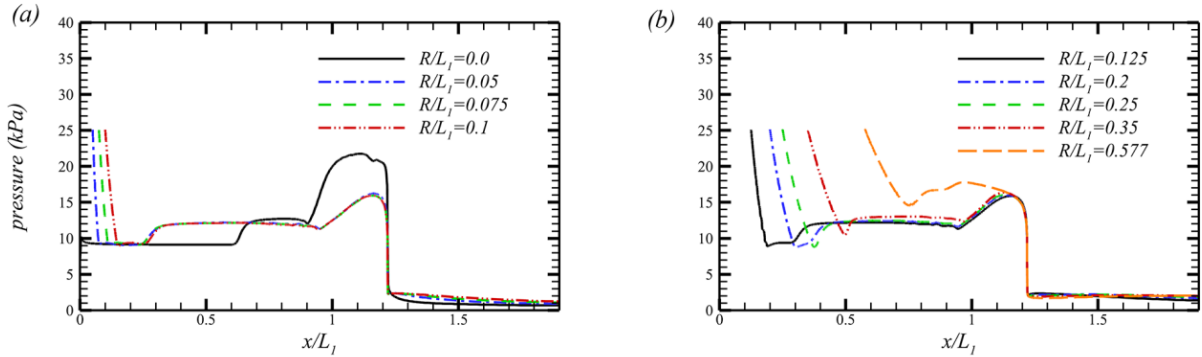

FIG. 11. Wall pressure distribution for different bluntness ratios at $t = 2.5$ ms for $\theta_2 = 45^0$, $L_1/L_2 = 2$ (a) $R/L_1 \leq 0.1$ (b) $R/L_1 > 0.1$

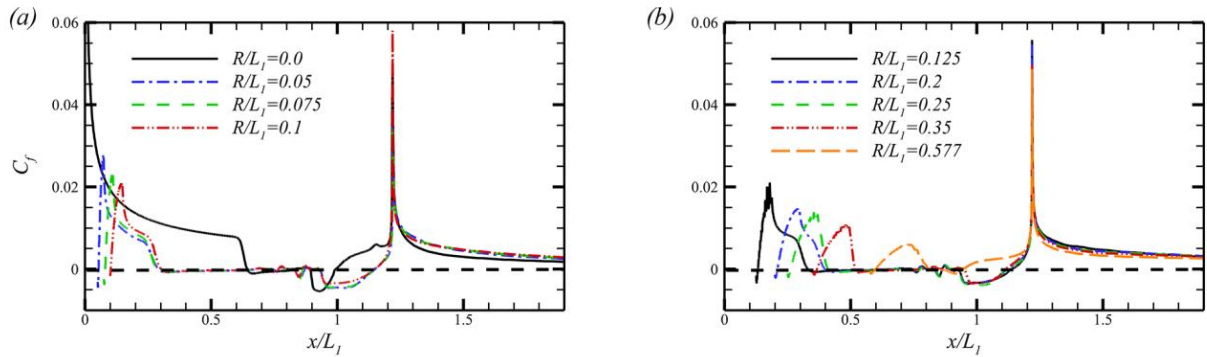

FIG. 12. Coefficient of skin friction distribution for different bluntness ratios at $t = 2.5$ ms for $\theta_2 = 45^0$, $L_1/L_2 = 2$ (a) $R/L_1 \leq 0.1$ (b) $R/L_1 > 0.1$

Figure 13 (a) provides the variation in separation location and re-attachment points for different bluntness ratios derived using skin friction distribution. These two points move far away from each other with the inclusion of a bluntness ratio of $R/L_1 = 0.05$. The re-attachment line remains nearly horizontal till $R/L_1 = 0.35$, after which it starts to decline. At the same time, the separation line moves upstream and gradually improves further from this location for the higher bluntness ratio configurations. Figure 13 (b) further shows the separation bubble length governed by the difference between these two locations, plotted against the bluntness ratios. It shows the sudden increase in the length of the separation bubble by a factor of *2.38* for the bluntness ratio of $R/L_1 = 0.05$ compared to the sharp double wedge case. It gradually peaks around $R/L_1 = 0.075$, indicating this ratio is an inversion bluntness ratio. These observations contrast the spherical bluntness added to the double cone configuration exposed to low enthalpy free stream condition, where the authors observe the separation bubble size remains unaltered until $R/L_1 = 0.075$ [30]. Moreover, they reported that the size of the separation bubble size grows gradually and peaks at a bluntness ratio between $R/L_1 = 0.075$ and $R/L_1 = 0.1$; its size then declines rapidly for higher bluntness ratio values. Here, the bubble size for $R/L_1 = 0.577$, in this case, is smaller than the sharp double wedge case, indicating the inversion radius in the range of $R/L_1 = 0.35$ and $R/L_1 = 0.577$.

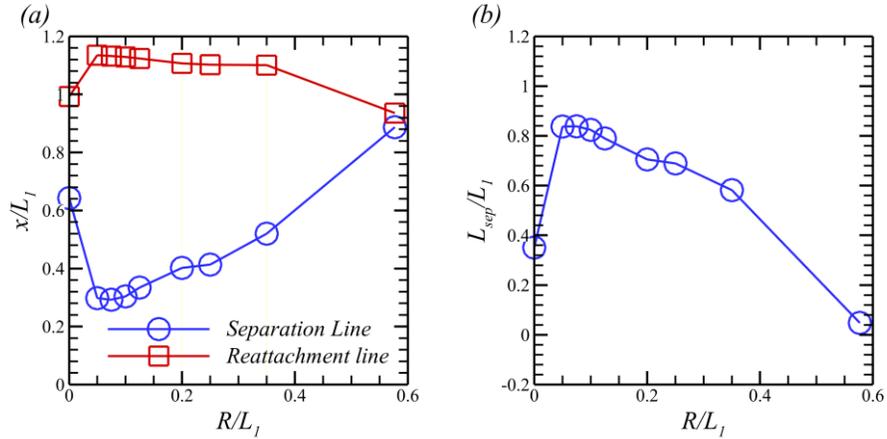

FIG. 13. (a) Variation of separation point and reattachment location (b) variation in the length of separation bubble as the function of bluntness ratios for the aft-wedge angle $\theta_2 = 45^0$

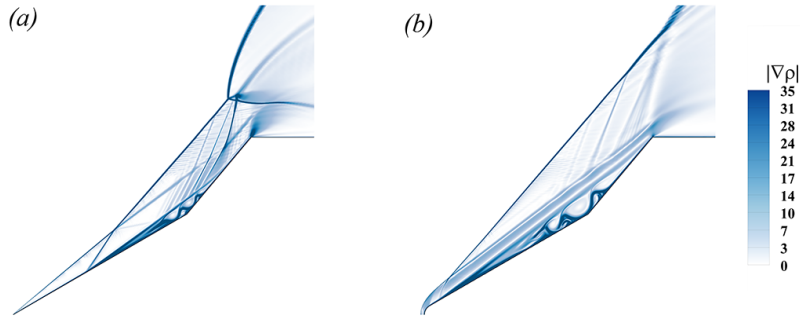

FIG. 14. (a) Comparison of the instantaneous numerical schlieren at $t = 2.5\ ms$ for $\theta_2 = 50^0$ (a) $R/L_1 = 0$ (b) $R/L_1 = 0.0$. Watch the animation for more details *(Multimedia view)*

## B. Aft-wedge angle $\theta_2 = 50^0$

The strength of the shock-wave boundary layer interaction is expected to increase with the increase in the aft-wedge angle, which was discussed intensively by Durna and Celik[12]. They observed a stronger type-V interaction for the sharp double-wedge with aft-wedge angle $\theta_2 = 50^0$, which generates a transmitted shock wave and impinges on the surface of the double-wedge. This was identified as the primary source of unsteadiness in the flow. At large flow-through times, the flow field sets itself in the self-sustained oscillatory state provided the transmitted shock impinges near the expansion corner. This was discussed in depth by Kumar and De[15] in their investigation of a different double wedge having varied combinations of the aft-wedge angles and wedge length ratios. Moreover, as observed in the previous section, the shock interaction pattern is altered to a great extent by including even a small amount of bluntness at the leading edge. In this framework, this work attempts to study the influence of leading-edge bluntness at higher aft-wedge angles.

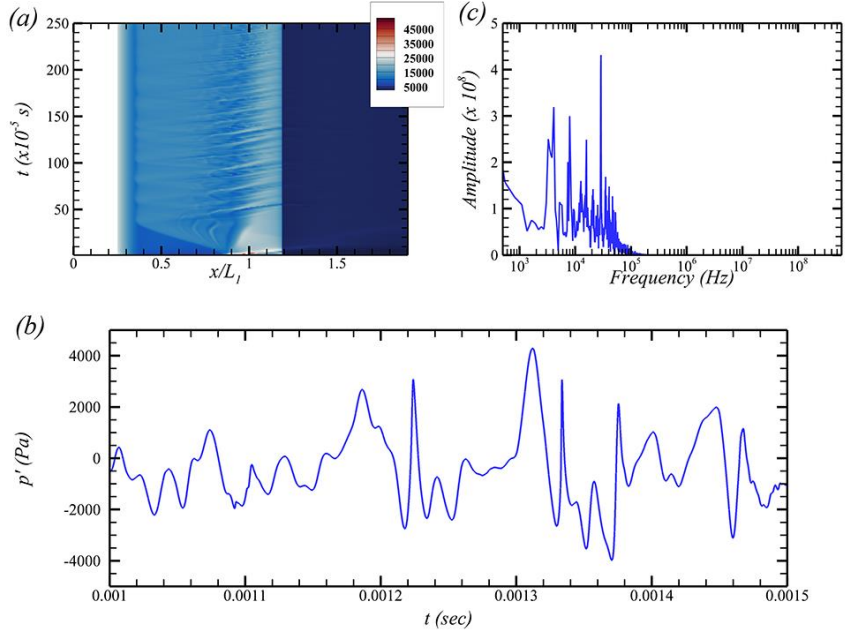

FIG. 15. (a) Spatial-temporal evolution of wall pressure coefficient. (b) Wall pressure signal at the compression corner ($x/L_1 = 0.866$) and (c) Fast Fourier Transform of the pressure fluctuations for $R/L_1 = 0.25$ and $\theta_2 = 50^0$

The aft-wedge angle is set at $\theta_2 = 50^0$, whereas the aft-wedge length is kept constant such that $L_1/L_2 = 2$ and add the bluntness ratios ranging in $0.0 \leq R/L_1 \leq 0.577$, similar to the previous case. Figure 14 (*Multimedia view*) represents the instantaneous flow field at $t = 2.5\ ms$ for $R/L_1 = 0$ and $R/L_1 = 0.05$. The flow features associated with sharp double-wedge configuration resemble Edney-type V interaction. They synchronize with the results published in the open literature[12-16], but the separation bubble elongates dramatically with the inclusion of a bluntness ratio as small as $R/L_1 = 0.05$ due to pressure over-expansion at the leading edge and undergoes Edney type-VI interaction. Along with this enlarged separation bubble, the shock structures undulate due to the rolling and interaction of the tertiary separation bubble formed beneath the primary separation bubble (flow field animation in the form of numerical schlieren and span wise vorticity is provided in Figure 14 (*Multimedia view*)). This case resembles the '*vibration mode*' earlier observed for the sharp double-wedge case with $\theta_2 = 55^0$ and $L_1/L_2 = 2$ [16]. But with the increase in the bluntness ratio, the flow undergoes unsteady oscillations for bluntness ratios $R/L_1 = 0.2$ onwards, unlike the previous case with the lower aft-wedge angle ($\theta_2 = 45^0$) and settles as a steady flow field at the highest possible bluntness ratio for this configuration, *i.e.* $R/L_1 = 0.577$. Note this is not the upper limit of the zone of the unsteady flow field, as simulations were not carried out exhaustively in this bluntness ratio range to determine the upper threshold limit. Figure 15 shows the spatial-temporal evolution of wall pressure along with the pressure fluctuation signal at the compression corner ($X/L_1 = 0.866$) and its *FFT* for one such unsteady case with $\theta_2 = 50^0$ and $R/L_1 = 0.25$. These figures uncover the presence of unsteadiness in the flow field with multiple dominant frequencies. This results from the increased strength of SBLI at this aft-wedge angle compared to the previous case, where a stronger interaction of Type V is expected during the initial development of the flow.[12]

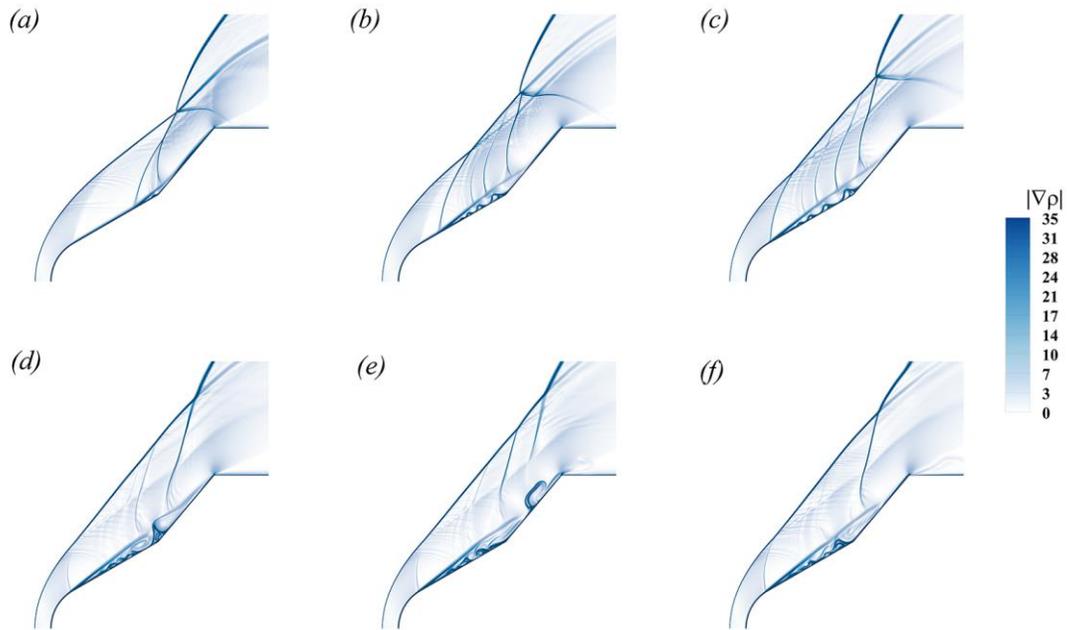

FIG. 16. Numerical Schlieren depicting the onset of unsteadiness at various arbitrary time instants during the initial flow development for $R/L_1 = 0.25$ and $\theta_2=50^0$ (a) $t = 110\ \mu s$. (b) $t = 250\ \mu s$ (c) $t = 320\ \mu s$ (d) $t = 490\ \mu s$ (e) $t = 530\ \mu s$ (f) $t = 570\ \mu s$. Watch the animation for more details *(Multimedia view)*

Figure 16 *(Multimedia view)* shows the cause of such unsteady interaction in the flow with the aid of numerical schlieren at arbitrary time steps but highlights the critical sequence of events for the case with $\theta_2 = 50^0$ and $R/L_1 = 0.25$. The flow field undergoes the Edney type-V interaction during the initial development stage. The shock interaction arising from the detached leading edge shock wave, the separation shock, and the bow shock wave generates an impinging shock wave of higher strength than the previous case (with the lower aft-wedge angle) and impinges on the aft-wedge surface. It thus offers the adverse pressure gradient (*APG*) to the flow field, and its disturbance propagates upstream via the boundary layer developing over the surface[12, 13]. Therefore, a large separation bubble is seen at the compression corner compared to the previous test case with the same leading edge bluntness ratio. The separation bubble grows under the transmitted shock's influence, pushing the separation

point upstream. The separation point crosses the point of tangency of the bluntness and the fore-wedge surface and leads to its drastic change in the direction at *t = 320 μs* in Figure 16 (c). The separation point continues to move upstream, following the contour of the blunted edge until it equilibrates with the upstream pressure. The sudden change in the separation point's movement acts as the source of unsteadiness in the flow and leads the shear layer to flap through which vortices shed, and it is explicitly shown in Figure 16 (e) (*t = 490 μs*). These shedding vortices interact with the secondary vortices beneath the shear layer and act in synergy to cause such a highly unsteady flow field with a spectrum of dominant frequencies. The animation for this case using numerical schlieren and span-wise vorticity contour is given as a multimedia file (Fig. 15 *(Multimedia view)*). Similar flow field behavior is observed at higher aft-wedge angles. The cause of such highly unsteady behavior will be explained further in this paper with the help of energy-based reduced order modeling, *i.e.* Proper Orthogonal Decomposition (*POD*).

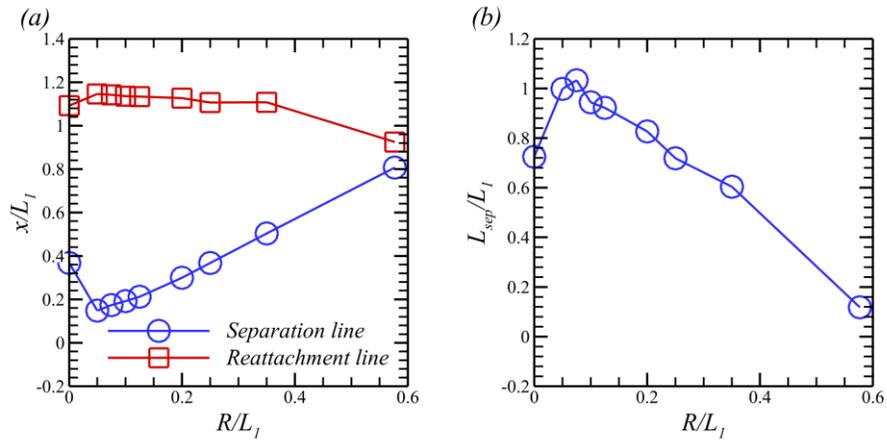

FIG. 17. (a) Variation of separation point and reattachment location (b) variation in the length of separation bubble as the function of bluntness ratios for the aft-wedge angle $\theta_2 = 50^0$

The investigation of determining the inversion and equivalent bluntness ratio is continued with the aid of the mean flow field despite the flow field being highly unsteady. 101 snapshots in a time averaging window of *t = 1 ms* and *2 ms* are utilized to calculate the mean. These bluntness ratios are derived using the mean wall shear stress distribution similar to the previous case ($\theta_2 = 45^0$). Figure 17 (a) shows the variation in the location of mean separation and re-attachment point. Their variation is similar to the previous case. However, the separation location and re-attachment move far apart, indicating the elongated separation bubble due to the increased interaction strength at the aft-wedge angle. Here also, one can notice a dramatic increase in the separation bubble size (Figure 17 (b)) with the inclusion of the blunted leading edge. The bubble size attains a global maximum at $R/L_1 = 0.075$ and declines thereafter with the increase in bluntness ratio. The separation bubble size is the least for $R/L_1 = 0.577$, significantly less than its sharp-double wedge contour part. The equivalent radius at this aft-wedge angle is approximately $R/L_1 = 0.25$ based on the mean separation bubble size. Not surprisingly, this portrays the shift in the equivalent radius to a lower bluntness ratio compared to the previous case (with a lower aft-wedge angle) owing to the increased strength of interaction at the higher aft-wedge angle, which ultimately increases separation bubble size for the sharp double wedge counterparts.

## C. Aft-wedge angle $\theta_2 = 55^0$

The aft-wedge angle is increased ($\theta_2$) further by $5^0$, maintaining the same aft-wedge length as the previous cases, and the bluntness ratios are added similarly. The flow field for the sharp double-wedge configuration at large flow through time is quasi-steady. It resembles the 'vibrational' mode of unsteadiness[16], where the shock structures oscillate about their mean position due to rolling the tertiary vortices. The separation bubble substantially elongates owing to the increased strength of the interaction of the impinging shock wave and the boundary layer compared to the previous double wedge configurations. Moreover, it elongates further with the addition of the leading edge bluntness, and this observation is consistent with the previous deductions for the lower aft-wedge angles. Additionally, the drastic shift in the interaction points of various shock waves emanating from different points is seen, which alters the shock interaction shape from strong Edney type V interaction to weak type VI shock interaction mechanism. Figure 18 shows the comparison of the two different flow fields at $t = 2.5\ ms$ depicting the essential features and highlighting the changes with the inclusion of the bluntness ratio. The observed differences are again attributed to the pressure over-expansion region by the blunted leading edge. However, as one can anticipate, the flow transits earlier at $R/L_1 = 0.1$ to an unsteady state in contrast to the previous case with a lower aft-wedge angle due to the increased bubble size arising out of the stronger interactions.

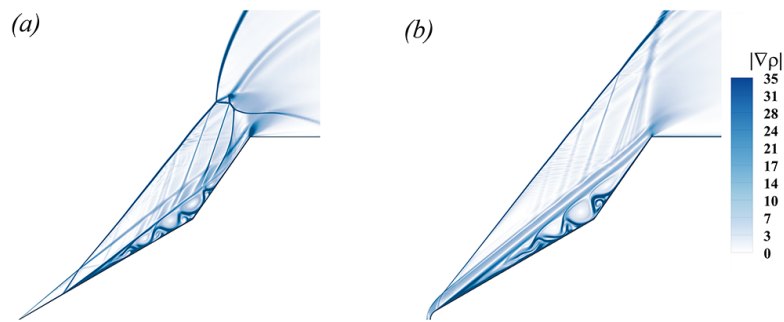

FIG. 18. (a) Comparison of the instantaneous numerical schlieren at $t = 2.5\ ms$ for $\theta_2 = 55^0$ (a) $R/L_1 = 0$ (b) $R/L_1 = 0.05$

Figure 19 (a) supports the observations for the unsteady flow at a higher bluntness ratio $R/L_1 = 0.25$ which shows a regular repeating pattern of the spatial-temporal distribution of the wall pressure at specific intervals for the case with aft-wedge angle set at $55^0$ and the bluntness ratio at $R/L_1 = 0.25$. Figure 19 (b) shows the wall pressure fluctuations acquired at the compression corner and *FFT* in Figure 19 (c). They reveal the first two dominating frequencies, $f_1 = 5450\ Hz$ and $f_2 = 9085\ Hz$, for the pressure fluctuation signal uncovering the important periodic phenomenon in the flow field. This case is emphasized as Swantek and Austin[8] used this sharp geometry counter-part in their experimental study to understand the impact of various free-stream flows using two different test gases. The flow field's modal decomposition is performed using the energy-based decomposition technique or Proper Orthogonal Decomposition (POD). Berkooz et al.[48] first utilized this technique to identify the presence of coherent structures in the turbulent flow field and has been used extensively in the fluid dynamics community[49-56] in recent times to identify the most energetic (most probable) structure in the flow field. Although this analysis tool shows the turbulent coherent structures, it has also been extensively used to identify the various modes of unsteadiness that might occur in the laminar flow field, such as laminar flow past vibrating and stationary slotted cylinders at laminar Reynolds number[57, 58]. In this context, this tool is utilized based on the algorithm suggested by Meyer et al.[59] to identify the cause of unsteadiness in the laminar flow field of the blunted double wedge configuration.

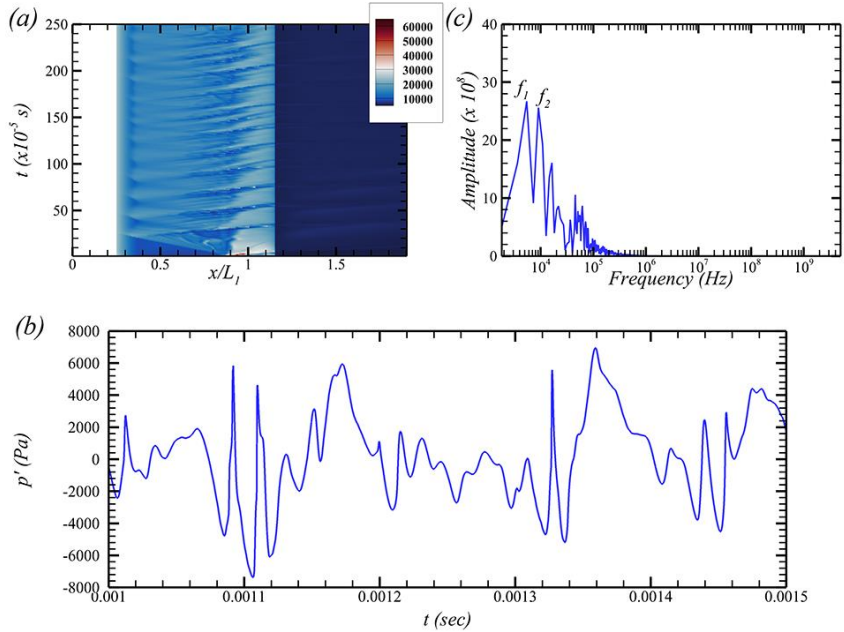

FIG. 19. (a) Spatial-temporal evolution of wall pressure. (b) Wall pressure signal at the compression corner ($x/L_1 = 0.866$) and (c) Fast Fourier Transform of the pressure fluctuations for $R/L_1 = 0.25$ and $\theta_2 = 55^0$ $f_1 = 5450\ Hz$ and $f_2 = 9085\ Hz$ (sampling frequency $f_s = 10,000\ MHz$)

The POD based on the method of snapshots requires an adequate number of flow field data to provide physically meaningful results. Therefore, a snapshot independence test is required to identify the most energetic structures in the flow field. Note the 'instantaneous snaps' of the flow field are utilized in this decomposition, where the first energetic Eigen basis function will correspond to the features of the mean flow field. It will give more precise insights into the mean flow field and separation bubble size. *3* snapshots are considered with *50, 100,* and *150* samples in the interval of $t = 1\ ms$ and $t = 2\ ms$ with the sampling rate of $\Delta t = 10\ \mu s$ and fixed initial time $t_i = 1\ ms$ and label them as set *1* through set *3*. Figure 20 (a) provides the singular values of the covariance matrix versus the number of snapshots for these *3* snapshots. These results show that the POD modes converge for the few modes but vary for very low energetic modes. However, there is minimal variation in energies (singular values) for the data set with *100* and *150* snapshots; therefore, the results for this analysis are reported with 100 snapshot sets. Figure 20 (b) elaborates on each Eigenmode's contribution to the total energy of the flow in the form of the cumulative contribution of the singular values for *100* set snapshots. As one can anticipate, the first mode corresponds to almost *98 %* of the total energy as the flow field is assumed to be laminar, containing all the mean structures of the flow field. The other modes do not contribute significantly to the total energy. Still, they represent essential structures that reveal the cause of unsteadiness and will be elaborated on further in this article. Also, note the singular values appear in mode pairs having comparable energies. This plot suggests that the first eight modes are sufficient to capture the flow field's total energy and can be used as the reduced order basis to reconstruct the entire flow field. These singular values are further shown in Table III, depicting the convergence for the various snapshots for the first eight energetic modes.

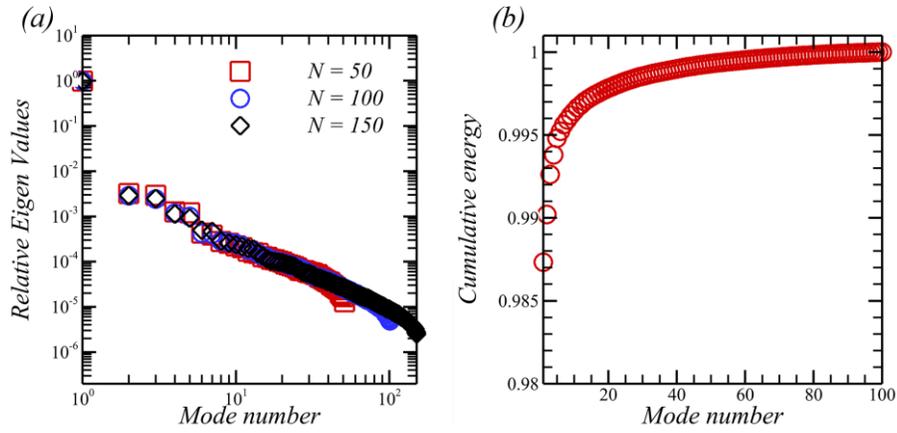

FIG. 20. (a) Singular values spectra for different sets of snapshots (b) Contribution of each energy mode to the total energy for 100 snapshot data for $R/L_1 = 0.25$ and $\theta_2 = 55^0$

TABLE III. Singular values for the most energetic modes with different sets of flow field snapshots.

| Mode No. | $\lambda_i$ | *N=50* | **N=100** | *N=150* |
|---|---|---|---|---|
| 1 | $\lambda_1$ | 0.98702 | **0.98733** | 0.98701 |
| 2 | $\lambda_2$ | 0.0032 | **0.0028** | 0.0028 |
| 3 | $\lambda_3$ | 0.0029 | **0.0024** | 0.0024 |
| 4 | $\lambda_4$ | 0.0012 | **0.0011** | 0.001 |
| 5 | $\lambda_5$ | 0.0012 | **0.0010** | 0.0009 |
| 6 | $\lambda_6$ | 0.0004 | **0.0004** | 0.0005 |
| 7 | $\lambda_7$ | 0.00038 | **0.0004** | 0.00046 |
| 8 | $\lambda_8$ | 0.00027 | **0.00028** | 0.00028 |

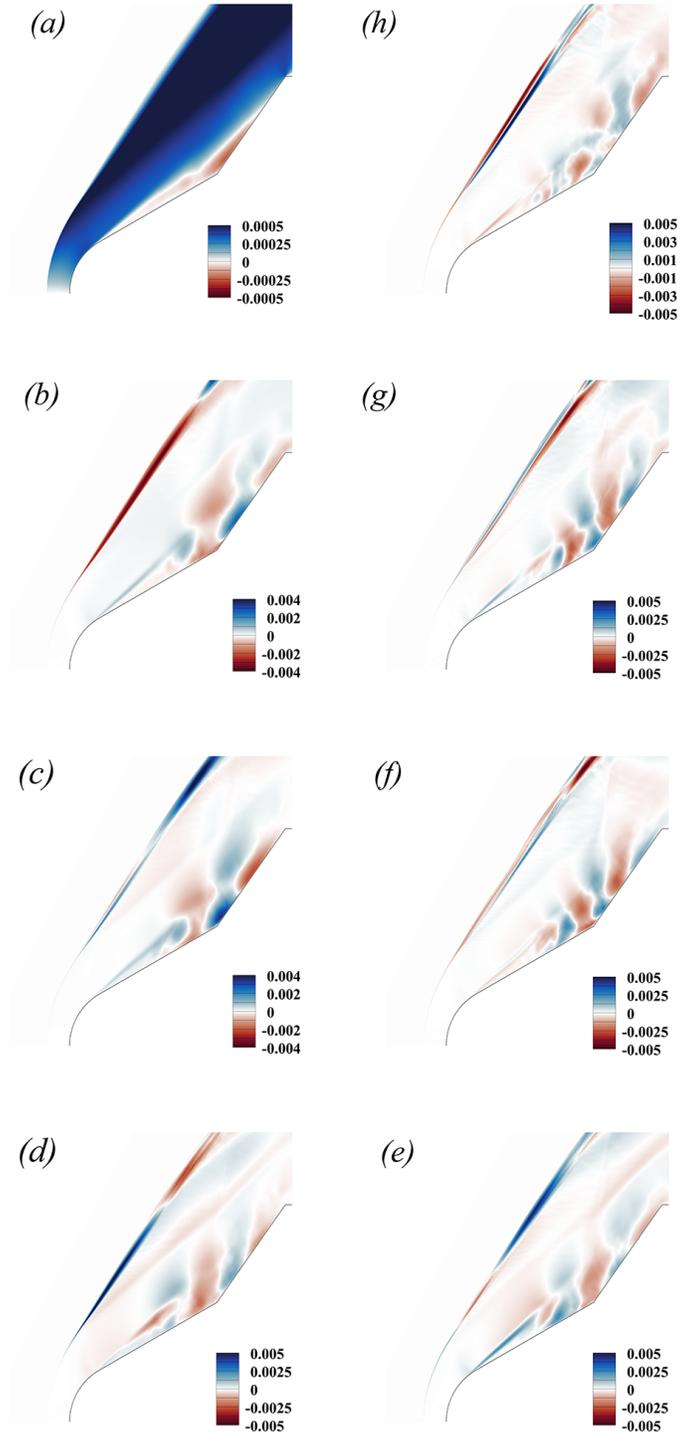

FIG. 21. First eight energetic Eigen function using 100 snapshots of the flow field in between $t = 1\ ms$ and $t = 2\ ms$ representation using the wall-normal component of the mode $\varphi_y$ for $R/L_1 = 0.25$ and $\theta_2=55^0$ (a) Mode 1 (b) Mode 2 (c) Mode 3 (d) Mode 4 (e) Mode 5 (f) Mode 6 (g) Mode 7 and (h) Mode 8

The two-dimensional energy-based POD contains 2 velocity components in the Cartesian coordinate system along with the temperature component. The energetic structures/modes are shown in Figure 21 using the wall-normal component $\varphi_y$. The most energetic structure, with almost *98.7 %* of the total energy shown in Figure 21 (a), resembles the mean flow field of the quasi-steady state cases (with low bluntness ratios). The second and third Eigen functions appear in mode pairs identified with a similar magnitude of the singular values, which hints at the convection phenomenon. Figures 21 (b) and 20 (c) depict the energies concentrated near the wall; however, they are shifted by a specific phase delay which ascertains the advection of these structures. They perhaps represent the convection of the shed vortices through the shear layer, which flaps due to instability generated by the separation point crossing the point of tangency, which was also observed for the lower aft-wedge angle case. The intermediate shock wave, too, appears with opposite energies, which once again confirms its oscillation as the traveling shock wave associated with the convection of vortical structures interferes with it periodically. Figure 21 (d) and 20 (e), the fourth and fifth most energetic mode shows the shear layer's flapping motion, originating from the boundary layer's separation point and anchors at the wedge's point tangency. The sixth and seventh modes shown in Figure 21 (f) and Figure 21 (g) have similar energy distributions and represent the small-scale structures convecting close to the wall. They greatly resemble the rolling of the small-scale tertiary vortices close to the point of tangency. However, these structures do not convect downstream and instead travel upstream. The least energetic mode 8 in Figure 21 (h) shows the randomly distributed energy structures near the wall; its contribution towards the flow's total energy is minimal.

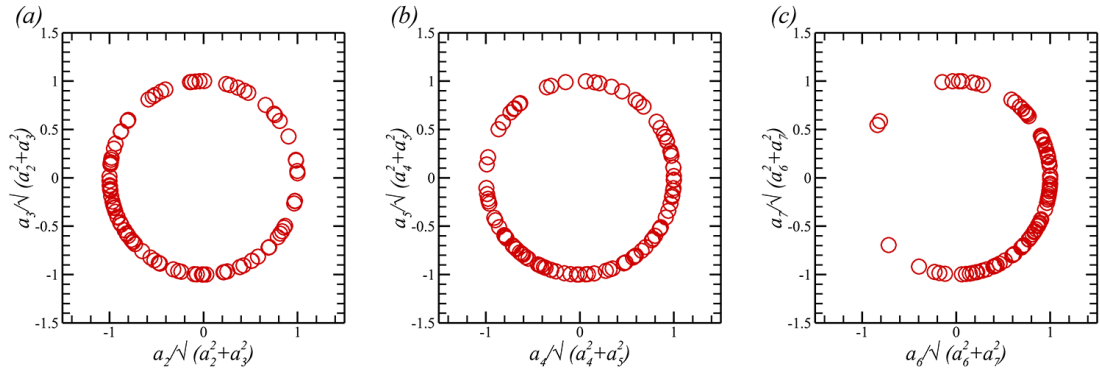

FIG. 22. Normalized temporal coefficients for the corresponding mode pairs (a) $a_2$-$a_3$ pair (b) $a_4$-$a_5$ pair (c) $a_6$-$a_7$ pair for *R/L₁ = 0.25* and *$\theta_2$=55⁰*

Figure 22 represents the temporal coefficient for the Eigenmode 2 - mode 3, mode 4 – mode 5, and mode 6 - mode 7 pairs in the form of phase portraits. Note the temporal coefficients are normalized concerning the amplitude in this plane to make them unit vectors. The coefficients are highly scattered along the unit circle but are clustered in some specific quadrants of the plane, which could be probably due to multiple frequency content in the decomposed time coefficient. This energy-based decomposition shows almost all the entire energies are concentrated in the mean flow along with the large-scale, small-scale convection and the perturbed shear layer and therefore explains the presence of a highly unsteady flow field. Nevertheless, the POD temporal coefficients varying with time and its corresponding FFT spectrum are also shown in Figure 23. The temporal coefficients for the corresponding mode pair vary similarly with time but are phase delayed. The FFT spectrum of these temporal coefficients also reveals the two dominating frequencies in close agreement with the frequencies obtained using the pressure signals at the compression corner shown in Figure 19. Note that this decomposition's sampling frequency is 100 kHz, which satisfies the Nyquist criteria to capture low frequencies of unsteadiness. This observation proves that the dominating energetic modes produced by POD strongly contribute to the unsteadiness in the flow field. Still, structures corresponding to the peak frequencies can only be identified using another Reduced Order Modelling approach known as the Dynamic Mode Decomposition (DMD) method. However, this is not in the scope of the present study.

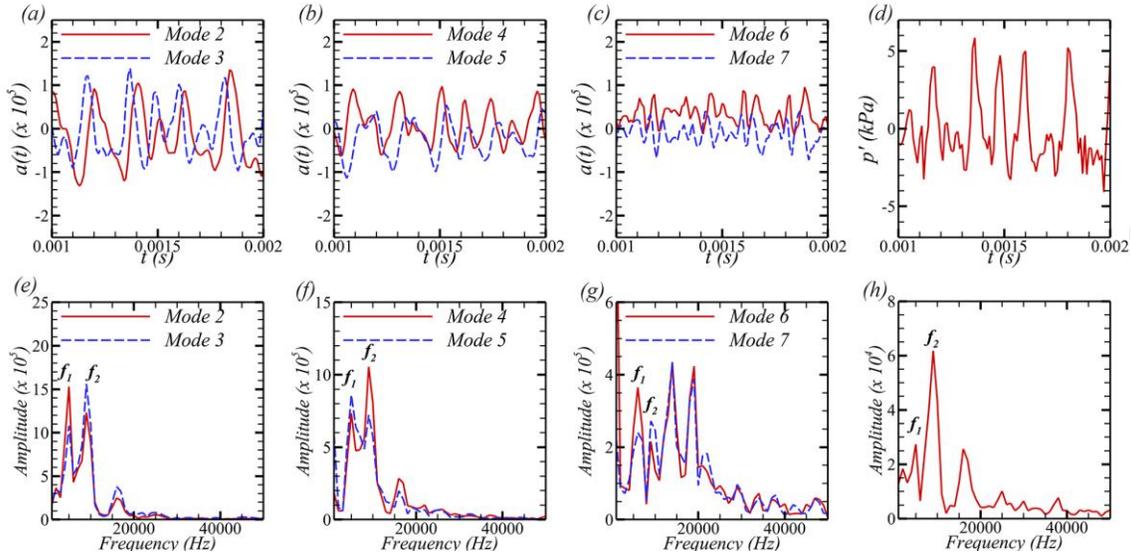

FIG. 23. Time variation of the temporal coefficients and its corresponding FFT.(a)-(c) time variation of the temporal coefficients of the POD modes (d) pressure fluctuation at the compression corner (e)-(g) FFT plots of the temporal coefficients of the POD modes (h) FFT of pressure fluctuation signal at the compression corner for $R/L_1 = 0.25$ and $\theta_2=55^0$. The pressure signal is down-sampled at 100 kHz. The solid red line shows the even mode's temporal coefficient and the blue dotted line shows the odd mode's temporal coefficient. Here, $f_1 = 5000\ Hz$ and $f_2 = 9000\ Hz$ (sampling frequency $f_s = 100\ kHz$)

Figure 24 reports the variation of locations of the point of mean separation and re-attachment with the bluntness ratios and the variation in the mean size of the separation bubble. These figures indicate the presence of an inversion bluntness ratio at $R/L_1 = 0.05$ and an equivalent bluntness ratio between $R/L_1 = 0.125$ and $R/L_1 = 0.2$. The inversion bluntness ratio for this case shifts to a lower value bluntness ratio compared to the previous lower aft-wedge angle cases. The reason for this shift is explained earlier in this article.

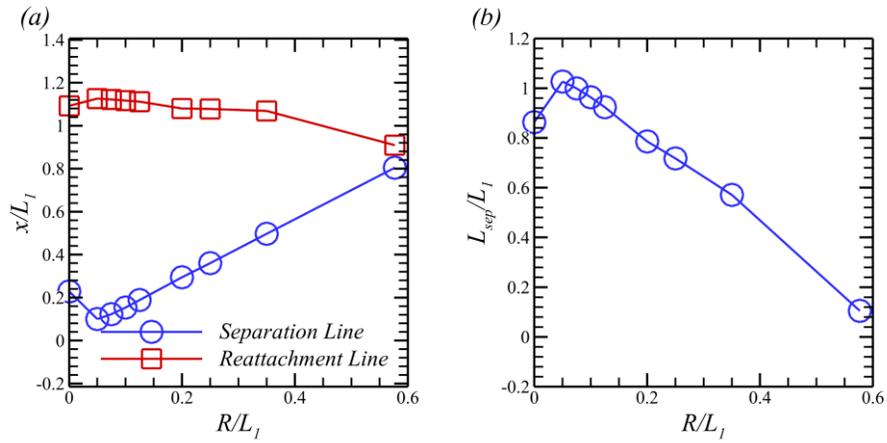

FIG. 24. (a) Variation of separation point and reattachment location (b) variation in the length of separation bubble as the function of bluntness ratios for the aft-wedge angle $\theta_2 = 55^0$

## D. Aft-wedge angle $\theta_2 = 60^0$

The aft-wedge angle is increased by $5^0$ further ($\theta_2$) to create stronger interaction and thus create a larger separation bubble size. The interaction mechanism changes entirely with the inclusion of the bluntness ratio and is in consistent agreement with the previous lower aft-wedge angle cases. Figure 25 presents these observations for the sharp double-wedge geometry and the double-wedge with leading edge bluntness ratio $R/L_1 = 0.05$. The flow field also transits to a highly unsteady state as well. Still, as one can anticipate, the transition of the flow field to an unsteady state occurs early at $R/L_1 = 0.075$ compared to the previous lower-aft wedge angles because of a considerable increase in SWBLI at this aft-wedge angle and separation bubble size.

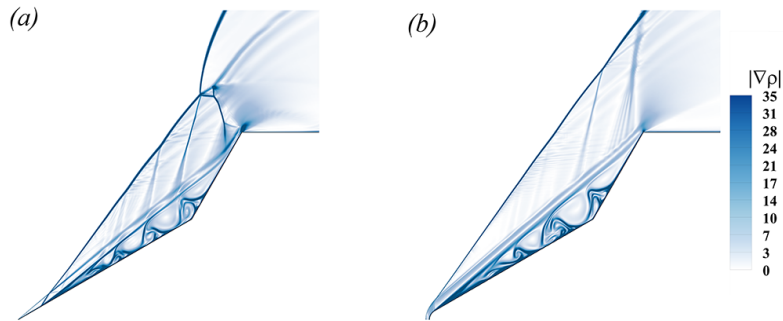

FIG. 25. (a) Comparison of the instantaneous numerical schlieren at $t = 2.5$ ms for $\theta_2 = 60^0$ (a) $R/L_1 = 0$ (b) $R/L_1 = 0.05$

Figure 26 further provides the spatial-temporal evolution of wall pressure distribution, the pressure fluctuation signals procured from the compression corner, and its *FFT* for $R/L_1 = 0.25$. The spatial-temporal evolution and pressure fluctuation signal in Figure 26 (a) and Figure 26 (b) respectively show the flow field to be highly unsteady with multiple dominant frequencies seen by its corresponding *FFT* plots, similar to the case with aft-wedge angle $\theta_2 = 55^0$ but, the frequencies are relatively lower compared to the previous cases. The cause of such unsteadiness is explained in the previous sub-section about the lower aft-wedge angle $\theta_2 = 50^0$ and $55^0$.

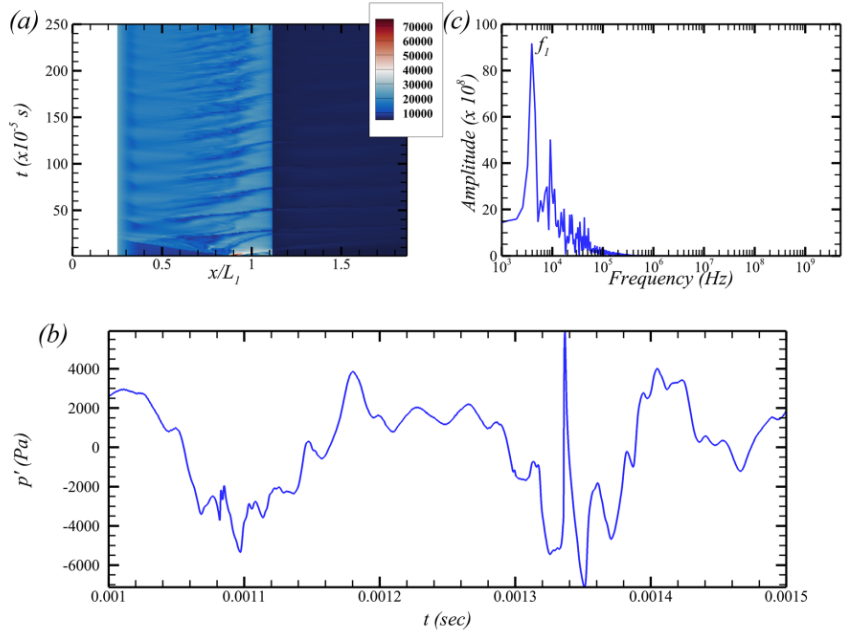

FIG. 26. (a) Spatial-temporal evolution of wall pressure. (b) Wall pressure signal at the compression corner ($x/L_1 = 0.866$) and (c) Fast Fourier Transform of the pressure fluctuations for $R/L_1 = 0.25$ and $\theta_2=60^0$. Here, $f_1 = 3980\ Hz$, $f_2 = 9145\ Hz$

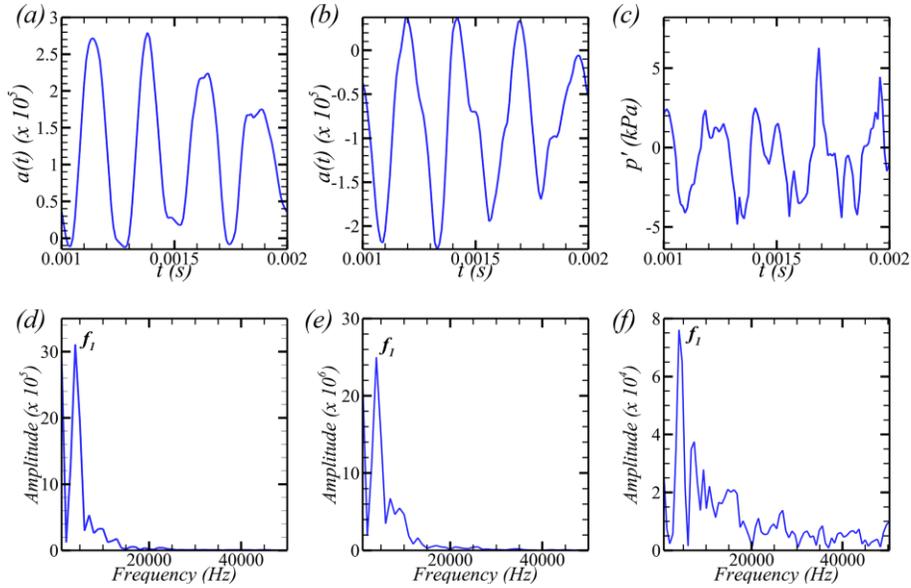

FIG. 27. Time variation of the temporal coefficients and its corresponding FFT. (a) mode 2 versus time (b) mode 3 versus time (c) pressure fluctuation at the compression corner (d) FFT of mode 2 (e) FFT of mode 3 temporal coefficient (f) FFT of pressure fluctuation signal at the compression corner for $R/L_1 = 0.25$ and $\theta_2=55^0$. The pressure signal is down-sampled at 100 kHz. Here, $f_1 = 4000\ Hz$ (sampling frequency $f_s = 100\ kHz$)

The flow field is decomposed further using *POD* to characterize the flow with similar arguments as was done earlier in this article. The Eigenmodes are not shown here for brevity, but Figure 27 reports the variation of temporal coefficients of the second and third Eigenmodes to analyze the frequency content of the flow. This figure indicates the presence of the first dominant at $f_1 = 4000\ Hz$ frequency, which is in close agreement with the frequencies revealed by the *FFT* of the pressure fluctuation signal in Figure 26. The second dominant frequency is suppressed in time coefficient frequency spectra. The contributing coherent structures at these frequencies can be confirmed with *DMD* analysis.

Figure 28 shows the mean location of re-attachment and separation points and mean separation bubble size as the function of bluntness ratios. The separation bubble size again increases drastically with the inclusion of leading edge bluntness at $R/L_1 = 0.05$. The separation bubble size peaks at $R/L_1 = 0.05$, equal to the inversion bluntness ratio obtained in the previous case. However, the equivalent bluntness ratio at this aft-wedge angle is approximately $R/L_1 = 0.075$, as the separation bubble size at this bluntness ratio matches approximately the case with a sharp leading edge.

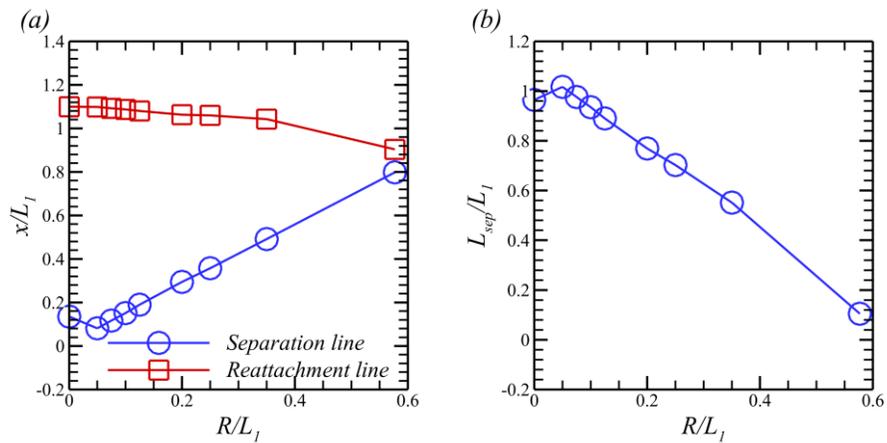

FIG. 28. (a) Variation of separation point and reattachment location (b) variation in separation bubble length as the function of bluntness ratios for the aft-wedge angle $\theta_2 = 60^0$.

## VII. SUMMARY

This article addresses the implication of the leading edge bluntness on the hypersonic flow over the double wedge with the multiple aft-wedge angle, covering various regimes of shock interference patterns. This study highlights the prominent differences in the flow structure and separation bubble size by including bluntness at the leading edge in contrast to the flow over the geometries with the sharp leading edge. Although its addition at the leading edge can be advantageous in terms of control over the separation bubble's size in weak interaction regions (type VI, weak type V) but such bluntness at a higher aft-wedge angle with strong shock interaction can lead to the undesirable flow field with unsteadiness, majorly arising due to the vortex shedding through the bluntness junction. Such unsteadiness may further amplify the existing instabilities in the system; hence, one must be careful while devising this kind of strategy to aid control.

The flow field for the low aft-wedge angle $\theta_2 = 45^0$ for all the bluntness ratios achieves a steady state at a sufficiently large flow-through time. The interference pattern for the sharp double-wedge at this aft-wedge angle resembles a weak Edney type-V interaction; however, the shock interference undergoes a drastic transformation with the detached leading edge shock and the curved shock waves. The separation bubble size also increases drastically by making the leading edge blunt. The peak value reaches the inversion ratio $R/L_1 = 0.075$ and declines further with the increase of the leading edge bluntness ratio. The equivalent ratio at this aft-wedge angle is between $R/L_1 = 0.35$ and $0.577$.

The aft-wedge angle is increased further by $5^0$, i.e. 50º, to increase the strength of the *SWBLI*. The interference pattern for sharp double-wedge is the Edney-V and is consistent with the observations in Ref 12, 13. The trends concerning the lower aft-wedge angle are also applicable for this aft-wedge case, but with bluntness ratio, $R/L_1 = 0.2$ onwards, the flow switches to the unsteady mode. The time history of wall pressure distribution, the pressure fluctuation signals at the compression corner, and the frequency transform for the double-wedge with bluntness ratio $R/L_1 = 0.25$ reveal a spectrum of dominant frequencies. The sudden change in the separation point's upstream movement perturbs the shear layer through which vortices shed and makes the flow-field unsteady. The quest to determine the critical bluntness ratios is carried out further, even though the flow field is unsteady using the mean flow field. The mean separation bubble size reaches the peak value at bluntness ratio $R/L_1 = 0.075$, and the equivalent bluntness ratio at this aft-wedge angle is approximately around $R/L_1 = 0.25$.

Further increasing the aft-wedge angle by $5^0$, i.e. 55º, remarkably increases the size of the separation bubble compared to the previous cases with lower aft-wedge angles (50º). The interference pattern for the sharp double-wedge configuration resembles the Edney type-V interaction. At large flow-through times, the shock structure vibrates about the mean position, which is facilitated by the rolling of the separation bubble. The flow, however, switches to the unsteady mode but at a lower bluntness ratio at $R/L_1 = 0.1$ compared to the previous case due to the increased strength of *SWBLI*. The *FFT* spectrum uncovers two dominant frequencies with $f_1 = 5450$ Hz, and $f_2 = 9085$ Hz, confirming the unsteady vortex shedding. The energy-based proper orthogonal decomposition reveals the coherent structures, and at the same time, the spectral analysis of the time coefficients also yields two dominating frequencies and agrees with the unsteady pressure fluctuations at the compression corner. The mean separation bubble at this aft-wedge angle reveals $R/L_1 = 0.05$ as the inversion bluntness ratio, and the equivalent ratio lies between $R/L_1 = 0.125$ and $R/L_1 = 0.2$.

This campaign extends to the highest aft-wedge angle, i.e. *60º*, and all the observations for the cases with lower aft-wedge angle are also true for this configuration. The flow transits to an unsteady state at a low bluntness ratio ($R/L_1 = 0.075$) in contrast to the previous cases. However, the spectral and POD analyses reveal unsteadiness frequency in the flow with $f_1 = 4000$ Hz, slightly lower than the previous case. The inversion and equivalent ratios are $R/L_1 = 0.05$ and $R/L_1 = 0.075$, respectively.

**VIII. CONCLUSIONS**

The present study deals with computational analysis of the double wedge with varied aft wedge angle and leading edge bluntness ratios. An improved solver, *'rhoCentralhypersonicRK4Foam'*, is validated using the experimental data of Swantek and Austin using a grid with the least error. The short time averaging window results are within the experimental uncertainty except near the peak heat flux location, and the agreement improves in the extended time-averaging window. The effects of several bluntness ratios are accessed using the validated solver for each aft-wedge angle by increasing it systematically from $\theta_2 = 45^0$ to $\theta_2 = 60^0$. The detailed analysis reveals the following outcomes:

- The impinging shock's strength increases with the increase in aft-wedge angle.

- The shock interaction changes with the inclusion of a small bluntness ratio ($R/L_1 = 0.05$) as compared to its sharp double-wedge counterpart.

- The mean separation bubble size increases drastically with the inclusion of blunted leading edge because of pressure over-expansion at the leading edge; it attains local maxima at the inversion bluntness ratio and declines further beyond the equivalent bluntness ratio.

- Both critical radii tend to shift to lower values of bluntness ratio with the increase in the aft-wedge angle (increase in the impinging shock strength).

- The flow field is steady for $\theta_2 = 45^0$ independent of the leading edge bluntness; however, bluntness ratios beyond this angle induce flow unsteadiness, and the impact is more predominant with the increase in the aft-wedge angle.

- Spectral analysis and proper orthogonal decomposition reveal the flapping of separated shear layers at the junction of bluntness addition through which vortices shed. The separation point could reach the bluntness junction owing to the high strength of impinging shock wave (transmitted shock), creating a larger separation bubble at higher aft-wedge angles.

On a closing note, the findings of the present study, such as the size of the separation bubble function of bluntness ratio and the flow unsteadiness, are based on the two-dimensional flow calculations. In reality, the actual double wedge model of Swantek and Austin is a full three dimensional model. There may be some three dimensional effect due to its small span-wise width-to-length ratio. The flow for such narrow geometry may experience side-ways relaxation (without lateral confinement) or spillage. This may create stream-wise streaks or Görtler vortices, introducing additional low frequency unsteadiness in the flow[60]. These aspects are not considered in the present analysis, and the current analysis results may slightly differ from the full three-dimensional calculations. Additional numerical calculations in direct numerical can provide deeper insights. Moreover, the present results are obtained for a particular free-stream condition. A scaling analysis may help to determine the impact of various free stream conditions on the unsteadiness and bubble size. This could be formulated as a universal law to quantify the effects of Mach number variation, stagnation enthalpy, and unit Reynolds number of the flow. Lastly, the present investigation utilizes POD to decode the most spatial probable structures, but DMD can effectively resolve the spatial-temporal coherent structures that can quantify the unsteadiness in the flow better. This article does not discuss these aspects due to their limited scope, and are left as future research work.

## ACKNOWLEDGMENTS

The authors would like to acknowledge the National Supercomputing Mission (NSM) for providing computing resources of "Param Sanganak" at IIT Kanpur, which is implemented by C-DAC and supported by the Ministry of Electronics and Information Technology (MeitY) and the Department of Science and Technology (DST), Government of India. We would like to acknowledge the IIT-K Computer Centre (www.iitk.ac.in/cc) for providing the resources to perform the computation work. This support is gratefully acknowledged.

## APPENDIX: MATHEMATICAL NOTATIONS

**Nomenclature**

| | |
|---|---|
| $L_1$ | Fore-wedge length |
| $L_2$ | Aft-wedge length |
| $\vec{W}$ | Vector of conserved variables |
| $\vec{F}_c$ | Convective flux vector |
| $\vec{F}_v$ | Viscous Flux vector |
| $\vec{Q}$ | Source Term vector |
| $E$ | Total energy of the flow $(J/kg)$ |
| $e$ | Internal energy $(J/kg)$ |
| $C_v$ | Specific heat at constant volume $(J/kgK)$ |
| $p$ | Thermodynamic Pressure $(N/m^2)$ |
| $I$ | Unit 2$^{nd}$ rank tensor |
| $T$ | Temperature $(K)$ |
| $\dot{q}_h$ | Heat generation per unit volume $(W/m^3)$ |
| $n$ | Unit normal vector of the control surface |
| $Pr$ | Prandtl Number |
| $R$ | Specific gas constant $(J/kgK)$ |
| $W$ | Arbitrary conserver variable |
| $n_f$ | Unit face normal vector of the control surface |
| $c$ | Speed of sound in the medium $(m/s)$ |
| $t_n$ | Time step at $n^{th}$ iteration level |
| $t_{n+1}$ | Time step at $(n+1)^{th}$ iteration level |
| $v_t$ | Component of velocity parallel to the patch |

**Greek Symbols**

| | |
|---|---|
| $\theta_1$ | Fore-wedge angle |
| $\theta_2$ | Aft-wedge angle |
| $\rho$ | Fluid's density $(kg/m^3)$ |
| $\bar{\bar{\tau}}$ | Viscous shear stress tensor $(N/m^2)$ |
| $\mu_v$ | Fluid's dynamic viscosity $(N\text{-}s/m^2)$ |
| $\Omega_i$ | Arbitrary control volume |
| $\lambda^+$ | Wave speed along the face normal $(m/s)$ |
| $\lambda^-$ | Wave speed opposite to the face normal $(m/s)$ |
| $\gamma$ | Ratio of specific heats of fluid |
| $\omega_f$ | Weighting factor |
| $\zeta(r)$ | Slope limiter function |
| $\alpha_l$ | Temporal integration weights for time integration |
| $\varphi$ | Arbitrary variable |
| $v_n$ | Component of velocity perpendicular to the patch |
| $U_\infty$ | Free stream velocity *(m/s)* |
| $p_\infty$ | Free stream pressure *(N/m$^2$)* |
| $T_\infty$ | Free stream temperature *(K)* |

## AUTHOR DECLARATIONS

### Conflict of Interest

The authors have no conflict of interest to disclose.

## DATA AVAILABILITY

The data that support the findings of this study are available from the corresponding author upon reasonable request.